\documentclass[rmp,aps,nofootinbib,twocolumn]{revtex4}
\usepackage{graphicx}
\usepackage{bm}
\usepackage{amssymb}

\def\inbar{\,\vrule height1.5ex width.4pt depth0pt}
\def\IR{\relax{\rm I\kern-.18em R}}
\def\IC{\relax\hbox{$\inbar\kern-.3em{\rm C}$}}

\def\bd{\hat\mathbf{ d}}
\def\bt{\mathbf{t}}
\def\beff{\mathbf{f}}
\def\bF{\mathbf{F}}
\def\bM{\mathbf{M}}
\def\da{\hat\mathbf{ d}_1}
\def\db{\hat \mathbf{d}_2}
\def\dc{\hat \mathbf{ d}_3}

\newcommand{\that}{\hat {\mathbf t}}
\newcommand{\nhat}{\hat {\mathbf n}}

\def\br{\mathbf{r}}
\newcommand{\bOm}{\mbox{\boldmath$\Omega$}}
\newcommand{\bom}{\mbox{\boldmath$\omega$}}
\newcommand{\brho}{\mbox{\boldmath$\rho$}}
\newcommand{\btheta}{\mbox{\boldmath$\theta$}}
\newcommand{\beps}{\mbox{\boldmath$\epsilon$}}
\newcommand{\bphihat}{\mbox{\boldmath$\hat\varphi$}}

\def\dd{\mathrm{d}}
\newcommand{\bnabla}{\mbox{\boldmath$\nabla$}}
\begin{document}
\title{Dynamics of filaments and membranes in a viscous fluid}

\author{Thomas R. Powers}
\email{Thomas_Powers@brown.edu}
\affiliation{Division of Engineering, Brown University, Providence, RI 02912, United States}

\begin{abstract}  
Motivated by the motion of biopolymers and membranes in solution, this article presents a formulation of the equations of motion for curves and surfaces in a viscous fluid. We focus on geometrical aspects and simple variational methods for calculating internal stresses and forces, and we derive the full nonlinear equations of motion. In the case of membranes, we pay particular attention to the  formulation of the equations of hydrodynamics on a curved, deforming surface. The formalism is illustrated by two simple case studies: (1) the twirling instability of straight elastic rod rotating in a viscous fluid, and (2) the pearling and buckling instabilities of a tubular liposome or polymersome. 
\end{abstract}                                                                 

\date{November 21, 2009}
\maketitle
\tableofcontents

\section{INTRODUCTION}
\label{sec:intro}

By definition,  soft matter is easy to deform. Some materials are soft because they have a high degree of symmetry. For example, the absence of long-range translational order in a 
liquid means that there is no resistance to shear in static equilibrium. Other materials, such as colloidal crystals, have low symmetry but are soft since the fundamental constituents are large and the interactions between these constituents are of entropic origin, and therefore weak. 
For example, to estimate the shear modulus $G$ of a colloidal crystal, we assume a characteristic length scale $a\approx1$\,$\mu$m and a characteristic energy scale of about ten times the thermal energy at room temperature, and find $G \approx 10k_\mathrm{B}T/a^3\approx4\cdot10^{-2}$\,Pa. This estimate is far smaller than the typical elastic modulus for a molecular crystal, where covalent bonds and molecular sizes determine the size of the shear modulus. Finally, many objects are soft since they are thin in one or two dimensions. It is a matter of common experience that bending a thin rod or plate is much easier than stretching. Examples of thin objects  at the micron scale abound: bacterial flagella, actin filaments, and cell membranes readily  deform in response to  thermal fluctuations or viscous forces arising in fluid flow. The key to understanding the physics of these systems often hinges on understanding the dynamics of their shapes. Thus, the student of soft matter is led directly to the problem of describing how curves and surfaces deform under external forces.  Differential geometry is the natural tool for this task. In this review, we give a self-contained introduction to differential geometry of curves and surfaces that evolve in time. Our intended audience is someone who wants to use these equations to elucidate physical phenomena.  Therefore, although our main focus is on mathematics,  we give physical interpretations of the mathematics whenever possible.
 
\begin{figure}
\centerline{%\centerbmp{4.95 in}{2.84 in}{hyperact-fig.bmp}
\includegraphics[height=2.5in]{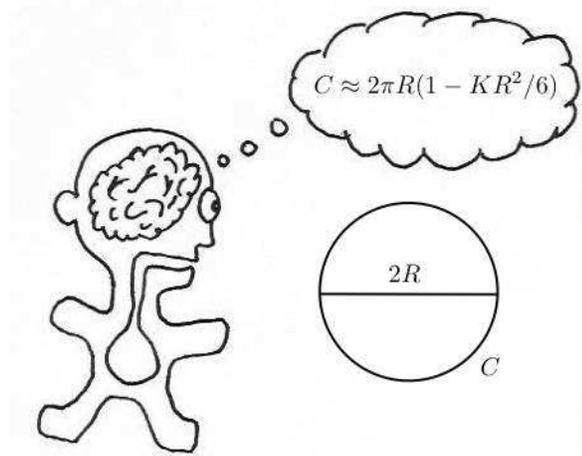}}
\caption{A two-dimensional being [after~\textcite{BurgerStarbird2005}]
considers the relation between the circumference and radius of a circle in his curved world; $K$ is a number he can determine by making careful measurements of length.}
\label{twoDcurvature}
\end{figure}

It is natural to ask why we need another review of differential geometry.  There are introductory books of a purely mathematical nature that are  excellent starting points for students
\cite{Coxeter1969,MillmanParker1977,struik1988,Morgan1998}. There are useful 
books with a broader mathematical scope, such as that of~\cite{Nakahara2003}. Finally, there are books and reviews that introduce differential geometry in the context of solid mechanics~\cite{GreenZerna1968}, fluid mechanics~\cite{Aris1989}, and soft matter~\cite{David1989,kamien2002,vanHemmenLeibold2007}. Our review is closest in spirit to these last four, but it is complementary since it emphasizes the variational point of view for deriving forces due to internal stresses acting on a stiff polymer or membrane. Like the review \cite{kamien2002},  we give a unified treatment of curves and surfaces from the ``extrinsic" point of view, in which distances along curves and surfaces sitting in three-dimensional space are measured using the notion of distance in that three-dimensional space. The alternative  ``intrinsic" approach to the differential geometry of surfaces is to work only with quantities that can be measured by a two-dimensional being living on the surface. For example, the curvature $K$ of a surface can be defined by comparing the circumference $C$ of a circle on the surface to the radius $R$ of the circle, as measured by the  inhabitant of the surface. (See Fig.~\ref{twoDcurvature}; the formula relating $K$, $C$, and $R$ displayed in the figure can be readily confirmed for the case of a sphere). Since polymers and membranes interact and bend in three-dimensional space, it is impossible to formulate their equations of motions in terms of intrinsic quantities only, and we will not hesitate to characterize curves and surfaces with extrinsic quantities such as length in three dimensions.

In addition to the tension between the extrinsic and intrinsic points of view, the issue of choosing coordinates always arises when studying curves and surfaces. Since geometric quantities such as vector fields and curvatures exist independently of the choice of coordinates, it is natural to define these quantities without invoking coordinates at all
~\cite{MisnerThorneWheeler}. Although this approach is elegant, it requires too much mathematical machinery for our purposes. 
In the case of curves, the basic geometrical ideas can be captured with vector calculus. Surfaces are more complicated and therefore require us to introduce some of the elements of the calculus of tensors, which we  attempt to do with a minimal amount of formalism.
 
A second reason not to insist on a coordinate-free approach is that  physical considerations often single out a coordinate system or a class of coordinate systems as special. For example, we will see that arclength parametrization is natural for inextensible rods, and  that convected coordinates---coordinates that are carried along by material particles---are natural for fluid membranes. A second point related to choice of coordinates is the distinction between an interface between two different phases and the physical surfaces that we study. For example, since a solidification front~\cite{Langer1980} advances through a substance while the material points of the substance remain at rest, we are free to use choose coordinates such that the interface velocity is always normal to the tangent plane~\cite{BowerKesslerKoplikLevine1984}.
In contrast, the velocity of a material point in a polymer or membrane may have a component along the tangential direction. For motion of a polymer or membrane arising from deformation, this velocity component has physical consequences and cannot be removed by a change of coordinates.

\section{CHARACTERISTIC LENGTH AND TIME SCALES}

\subsection{Thermal fluctuations and the persistence length}
\label{ThermalFlucts}

To keep the review of manageable size, we limit the scope by disregarding thermal fluctuations. 
This assumption is a vast simplification, since the flexibility of polymers and membranes ensures that thermal undulations are always present. 
However, there are many shape-evolution problems in which these effects are small.
Furthermore, the formalism described in this review is also useful even for  problems where thermal effects play an important role, such as the dynamics of tension in semiflexible polymers~\cite{HallatschekFreyKroy2007} and the surface tension 
of an undulating fluid membrane~\cite{FournierBarbetta2008}. In this section we briefly delineate the conditions under which thermal fluctuations are important.

First consider a filament in solution. The filament could be a semiflexible polymer such as DNA or actin, or the flagellar filament of a bacterium~\cite{boal2002}. These filaments resist bending, and may be described by an energy 
\begin{equation}
E=\frac{A}{2}\int\frac{1}{R^2}\ \dd s,
\end{equation}
where $A$ is the bending modulus, with dimensions of energy times length, and $s$ is distance measured along the filament, and $R$ is the local radius of curvature of the filament. We will describe the curvature of a filament in much greater detail in \S\ref{sec:RodKinematics}; for now the reader can simply imagine that the thermal motion of the molecules of the solution causes the filament to flex back and forth between arcs of constant curvature. If we imagine lengthening the filament, then our experience with macroscopic rods tells us that the filament is easier to bend. That is, for fixed filament stiffness $A$, the same force applied perpendicular to the end of the filament leads to a greater deflection as the filament gets longer. 
Thus, the longer the filament, the more it will bend spontaneously due to  thermal fluctuations.  We can estimate the length $\ell_\mathrm{p}$ of a filament at which thermal fluctuations become important by equating thermal energy $k_\mathrm{B}T$ with the bending energy of a filament bent into an arc of one radian: $k_\mathrm{B}T\sim A/\ell_\mathrm{p}$, or $\ell_\mathrm{p}\sim k_\mathrm{B}T/A$. The length $\ell_\mathrm{p}$ is called the persistence length. At room temperature, the persistence length of DNA is about 50\,nm; $\ell_\mathrm{p}$ is of the order of microns for actin, and millimeters for microtubules and bacterial flagellar filaments~\cite{boal2002}.  The effects of thermal fluctuations are small when the length $L$ of a filament is small compared to $\ell_\mathrm{p}$, $L\ll\ell_\mathrm{p}$. 

Unlike interfaces which are simply the boundary between two phases, membranes are comprised of molecules that resist bending. Therefore, the shapes of membranes are also governed by a bending energy. As we review in \S\ref{sec:GeomSurface}, there are two radii of curvature at any point of a surface. Thus, the bending energy of a fluid membrane is
\begin{equation}
E=\frac{\kappa}{2}\int\left(\frac{1}{R_1}+\frac{1}{R_2}\right)^2\dd S,\label{membenden}
\end{equation}
where $\kappa$ is the bending modulus or stiffness, with dimensions of energy, and $\dd S$ is the element of of surface area. Just as for filaments, we expect that as a membrane gets larger and larger, the thermal motion of the solvent is more and more effective at bending the membrane. However, since the units of $\kappa$ do not contain a length, it is not immediately obvious how to estimate the persistence length using dimensional analysis. Therefore, we must use a more detailed approach.

Consider a  membrane that is weakly perturbed by thermal motion and spanning a frame with dimensions $L\times L$. Since the membrane is almost flat, it can be described as the graph of a height function $h$ over the $xy$-plane. In \S\ref{sec:GeomSurface} we will see that the energy in Eq.~(\ref{membenden}) may be approximated as 
\begin{equation}
E\approx\frac{\kappa}{2}\int(\nabla^2h)^2\dd x\dd y.\label{Equadratic}
\end{equation}
Our strategy for quantifying the effect of thermal fluctuations will be to estimate the scale $L$ at which fluctuations in the normal to the membrane, $\nhat\approx\hat\mathbf{z}-\bnabla h$, become of the order of unity, or in other words, large enough to spoil the assumption that the membrane is almost flat. First, use the equipartition theorem to assign to each mode of the quadratic energy function (\ref{Equadratic}) the contribution $k_\mathrm{B}T/2$ toward the average energy. The simplest way to apply the equipartition theorem is to assume periodic boundary conditions and write the height as a Fourier series, $h(x,y)=\sum_\mathbf{q}\exp[\mathrm{i}\mathbf{q}\cdot(x,y)]h_\mathbf{q}/L^2$, where the sum is over all wavevectors $\mathbf{q}=(2\pi m/L,2\pi n/L)$ with the integers $m$ and $n$ less than a maximum value $m_\mathrm{max}$. The wavelength corresponding to the maximum is the smallest length at which the continuum theory applies, and is a molecular scale $\ell_\mathrm{mol}$ such as the thickness of the membrane. Thus, the mean-square amplitude of the fluctuations of the height are 
\begin{equation}
\langle h_\mathbf{q}h_{\mathbf{q}'}\rangle=L^2\delta_{\mathbf{q}+\mathbf{q}',\mathbf{0}}\frac{k_\mathrm{B}T}{q^4}.\label{hamp}
\end{equation}
Longer wavelength ripples are easier to excite. We can now calculate the fluctuations in the normal direction $\nhat$. Since the membrane is almost flat, it is convenient to calculate the mean-square fluctuations of the deviation $\delta\mathbf{n}=\nhat -\hat\mathbf{z}\approx-\bnabla h$~\cite{ChaikinLubensky1995}:
\begin{eqnarray}
\langle\delta\nhat\cdot\delta\nhat\rangle&=&\sum_{\mathbf{q}\mathbf{q}'}\frac{\mathbf{q}\cdot\mathbf{q}'}{L^2}\langle h_\mathbf{q}h_{\mathbf{q}'}\rangle\\
&\approx&\int\frac{\dd^2q}{(2\pi)^2}\frac{k_\mathrm{B}T}{\kappa q^2}\label{qint}.
\end{eqnarray}
The integral in Eq.~(\ref{qint}) runs over wavevectors ranging from the short-distance cutoff $q_\mathrm{min}\approx2\pi/\ell_\mathrm{mol}$ to the long-distance cutoff given by the scale of the frame, $q_\mathrm{max}\approx 2\pi/L$. Assuming for simplicity a disc-shaped domain of integration, we find 
$\langle\delta\mathbf{n}\cdot\delta\mathbf{n}\rangle\approx k_\mathrm{B}T/(2\pi\kappa)\log\left(L/\ell_\mathrm{mol}\right)$. The scale at which the thermal fluctuations lead to significant deformation of the membrane is the length scale $L=\ell_\mathrm{p}$ for which $\langle\delta\mathbf{n}\cdot\delta\mathbf{n}\rangle\approx1$, or
\begin{equation}
\ell_\mathrm{p}\approx \ell_\mathrm{mol}\exp\left(\frac{2\pi\kappa}{k_\mathrm{B}T}\right).
\end{equation}
For typical values $\ell_\mathrm{mol}\approx 1$\,nm and $\kappa\approx10k_\mathrm{B}T$, the persistence length is extremely large. Thus, thermal fluctuations are often disregarded in studies of the shape dynamics of vesicles.

\subsection{Time scales for stretching, bending, and twisting of filaments}\label{sec:TimeFilament}
To understand the physical mechanisms that govern the dynamics of filaments, we describe the characteristic time scales for stretching, bending, and twisting. To get the orders of magnitude of these time scales, it is sufficient to use scaling arguments and work in the limit of small deflection from a straight filament. The precise nonlinear equations of motion require the development of the geometric tools presented in the later sections of this review. 

We begin by recalling some basic definitions from the mechanics of rods~\cite{landau_lifshitz_elas} and viscous fluids~\cite{landauFM}. Consider a straight rod of radius $a$, with long axis aligned along $z$ and under longitudinal deformation. The stress $\sigma_{zz}$, or force per unit area acting through a cross-section at $z$, is proportional to the strain for small deformations: $\sigma_{zz}=Y\partial_z u$, where $u$ is the displacement of the cross-section of the rod originally at $z$, and $Y$ is the Young's modulus with units of energy per volume. The net elastic force per length acting on the element of size $\dd z$ is therefore given by $\pi a^2Y\partial^2_z u\ \dd z$. Now suppose the rod is allowed to relax, with the only resistance to the relaxation being the viscous drag of the solvent. 

To estimate the viscous resistance, we first argue that at the small scales of polymer and cells, the typical viscous forces are much larger than the forces that are required to change the velocity of the fluid. Viscous stresses are given by the product of the viscosity $\eta$ and the strain rate. Thus, the viscous force on an object of size $\ell$ is approximately 
\begin{equation}
F_\mathrm{v}\approx(\eta v/\ell)\ell^2=\eta v\ell,
\end{equation}
where $\eta$ is the viscosity and $v$ is a characteristic speed of the flow. The viscosity of water is $\eta\approx10^{-3}$\,N-s/m$^2$.
The product of mass and acceleration for an element of size $\ell$ of a fluid of density $\rho$ is approximately
\begin{equation}
F_\mathrm{i}\approx(\rho \ell^3)(v^2/\ell)=\rho v^2 \ell^2.
\end{equation}
For micron-sized objects in water moving at speeds of the order of microns per second, the inertial force $F_\mathrm{i}$ is about a million times smaller than the viscous force $F_\mathrm{v}$. This regime is the regime of ``low Reynolds number," in which inertia is irrelevant~\cite{happel,kimbook,hinch_lowRe}. Thus, we may estimate the viscous drag on the element of the rod of length $\dd z$ to be of the order of $\eta v\dd z$. (In \S\ref{sec:curves} we address the question of how the length scale $a$ enters the drag.) 

We now have the ingredients required to estimate the relaxation time for longitudinal extension and compression of a filament. Assuming that the inertia of the rod as well as the fluid is negligible, we balance the viscous drag with the elastic force, identify $v$ with $\partial_t u$,  and drop numerical factors such as $\pi$ to find 
$\eta\partial_t u\sim a^2 Y\partial_z^2 u$.
Thus, the characteristic relaxation time for longitudinal stretching or compression of a rod of length $L$  in a viscous fluid is
\begin{equation}
\tau_\mathrm{s}\sim \frac{\eta}{Y} \left(\frac{L}{a}\right)^2.
\end{equation}

To estimate the relaxation time for bending deformations, note that the bending modulus $A$ for a thin filament of radius $a$ is related to the Young's modulus by $A\sim Ya^4$~\cite{landau_lifshitz_elas}. Just as we did for membranes, we can describe small deflections of a filament by a height function $h(z)$, and approximate the elastic energy as $E\approx(A/2)\int(\partial_z^2h)^2\dd z$. By the principle of virtual work, the elastic force per unit length due to bending is given by the functional derivative $-\delta E/\delta h\sim-A\partial_z^4 h$. Once again disregarding numerical factors, we approximate the resistance to transverse motion as $\eta\partial_t h\dd z$, and balance the elastic and viscous forces to obtain the bending relaxation time
\begin{equation}
\tau_\mathrm{b}\sim\frac{\eta L^4}{A}\sim\frac{\eta}{Y}\left(\frac{L}{a}\right)^4.
\end{equation}
The ratio of the stretch relaxation time to the bend relaxation time involves two powers of the aspect ratio, $\tau_\mathrm{s}/\tau_\mathrm{b}\sim(a/L)^2$. This aspect ratio is typically very small, ranging from $1/50$ for a 50\,nm segment of DNA to $10^{-3}$ for a typical bacterial flagellum~\cite{boal2002}. Therefore, modulations in the extension relax much faster than bending deformations for polymer filaments. When the focus of interest is bending dynamics, thin filaments are typically assumed to be inextensible

To estimate the relaxation time for twisting deformations, suppose the filament is straight but twisted. We will precisely define twist for a filament with arbitrary contour in \S\ref{sec:RodKinematics}, but for a straight filament define the angle $\theta(z)$ through which the cross-section at $z$ has rotated due to twist. The torque applied through a cross-section is $C\partial_z\theta$, and the net elastic torque on an element of length $\dd z$ due to internal stresses is $C\partial_z^2\theta\dd z$. 

The viscous torque resisting rotation at rate $\partial_t\theta$ is the product of the viscous stress $\eta\partial_t\theta$, the surface area of the element $\sim a\dd z$, and the moment arm $a$. Thus, the twist degree of freedom obeys the diffusion equation $\eta a^2\partial_t\theta=C\partial_z^2\theta$, and the twist relaxation time is
\begin{equation}
\tau_\mathrm{t}\sim\frac{\eta a^2L^2}{C}.
\end{equation}
For many materials, $C\sim A$~\cite{landau_lifshitz_elas}, and thus
\begin{equation}
\tau_\mathrm{t}\sim\frac{\eta}{Y}\left(\frac{L}{a}\right)^2.
\end{equation}
The twist relaxation time is much smaller than the bend relaxation time. Interestingly, the twist relaxation time is comparable to the stretch relaxation time. Nevertheless, it is common to consider the dynamics of inextensible elastic rods with both bending and twisting degrees of freedom, since twisting and stretching are usually driven by different physical processes. A simple example is provided by an elastic filament, initially straight and twirling about its long axis (\S\ref{illustrative1}). Since the rotation induces twist but no stretching, it is convenient to treat the filament as inextensible.

\subsection{Time scales for stretching and bending of membranes}
In this section we review the basic mechanics of membranes~\cite{seifert1997}.
We confine our attention to fluid membranes, in which there is no long-range order in the molecules making up the membrane. These membranes wrap up into closed surfaces known as vesicles---liposomes in the case of lipid molecules, and polymersomes in the case of block copolymers~\cite{DischerAhmed2006,YuEisenberg1998}. Both the lipids and the block copolymers are amphiphilic, consisting of two parts with different affinities for water. The hydrophobic effect drives the molecules to form a bilayer, with the oily chains with the least affinity for water inside the bilayer. 

We now estimate the relaxation time for membrane stretching. Consistent with the discussion of \S\ref{ThermalFlucts}, we suppose the membrane is in the ``high-tension" regime of stretching in which the resistance arises from elasticity rather than entropic effects~\cite{EvansRawicz1990}.  Disregarding the bilayer structure for the moment, we may view the membrane as a two-dimensional fluid. 
Unlike the filaments we considered in \S\ref{sec:TimeFilament}, the membrane is a liquid, and to describe the resistance of the membrane to stretching and compression we use the density rather than a displacement variable. Let $\phi$ denote the number of molecules per unit area, and $\phi_0$ denote the preferred number density. For small deformations, the elastic energy of stretching is
\begin{equation}
E_\mathrm{s}=\frac{k}{2}\int\left(\frac{\phi}{\phi_0}-1\right)^2\dd S,
\end{equation}
where $k$ is the compression modulus with units of energy. Since the modulus $k$ is usually large enough that the density is close to $\phi_0$, work in terms of $\delta\phi=\phi-\phi_0$. From dimensional analysis, we deduce that the two-dimensional pressure in the membrane, with units of force per length, is $p_{2\mathrm{d}}=k\delta\phi/\phi_0$. Gradients in the two-dimensional pressure lead to tangential forces in the membrane, which drive regions of non-uniform density to relax to constant density. The balance of these tangential forces with dissipative forces determines the stretch relaxation time. We consider the relaxation of a density modulation with wavenumber $q$. 

First we suppose that the dominant dissipative forces arise from the viscous forces the solvent exerts on the membrane. For a characteristic flow field $v$ near the membrane, the stress is $\sim\eta\nabla v\sim\eta q v$. Thus the balance of the gradient of pressure with the viscous stress leads to 
\begin{equation}
qk\delta\phi/\phi_0\sim\eta q v.\label{gpresvstress}
\end{equation}
 Assuming the no-slip boundary condition between the solvent velocity and the lipid velocity holds, we may use $v$ to represent the lipid velocity as well. The rate of change of the density $\phi$ is related to the lipid flow by the continuity equation expressing conservation of particles. In \S\ref{sec:DynamicalEquationsMembrane} we will see how to formulate the continuity equation on a moving curved surface, but for our scaling argument it is sufficient to assert that $\partial_t\phi\sim\nabla(\phi v)$, or 
\begin{equation}
\delta\phi/\tau_\mathrm{s}\sim q\phi_0 v,\label{lipconsscaling}
\end{equation}
where $\tau_\mathrm{s}$ is the stretching relaxation time of the membrane.  Combining Eqs.~(\ref{gpresvstress}) and (\ref{lipconsscaling}) leads to 
\begin{equation}
\tau_\mathrm{s}\sim\frac{\eta}{kq},
\end{equation} or, taking the smallest wavenumber to be set by the size $L$ of the vesicle, 
$\tau_\mathrm{s}\sim\eta L/k$.

To estimate the relaxation time for bending deformations with wavenumber $q$, we balance the bending elastic force per unit area with viscous stress or traction from the solvent. We will give the precise form of the elastic force per area on a deformed membrane in \S\ref{sec:fluidmembrane}, but working in analogy with our expression for the bending force per unit length on a filament, we have $f\sim\kappa\nabla^4 h\sim\kappa q^4h$. Again invoking the no-slip condition to identify $\partial_t h$ with the velocity in the solvent near the membrane, we estimate the viscous traction on the membrane as $\eta q h/\tau_\mathrm{b}$, where $\tau_\mathrm{b}$ is the bending relaxation time scale. Balancing the elastic force per area with the traction leads to~\cite{BrochardLennon1975}
\begin{equation}
\tau_\mathrm{b}\sim\frac{\eta}{\kappa q^3}.\label{bendrelaxtime}
\end{equation}
Since the smallest $q$ scales with vesicle size $L$, we may also write $\tau_\mathrm{b}\sim\eta L^3/\kappa$. A simple scaling estimate for the magnitude of the bending stiffness is $\kappa\sim k\ell_\mathrm{mol}^2$, where $\ell_\mathrm{mol}$ is a molecular scale such as the thickness of the membrane~\cite{seifert1997}. Thus, the ratio of the stretch relaxation time to the bend relaxation time for membranes scale as $\tau_\mathrm{s}/\tau_\mathrm{b}\sim(\ell_\mathrm{mol}/L)^2$, and just as in the case of filaments we expect density modulations to relax much faster than bending modulations. To estimate the time scales directly from the measured parameters for the lipid stearoyloleoyl phosphatidylcholine (SOPC), we use $k\approx0.2$\,N/m and $\kappa\approx10^{-20}$\,J~\cite{Dimova_etal2002}, which yields for a vesicle with $L=10$\,$\mu$m a stretch relaxation time of $\tau_\mathrm{s}\approx10^{-7}$\,s and a bending relaxation time of $\tau_\mathrm{}\approx100$\,s.~\footnote{In principle, the relaxation dynamics of the compression modes could be so fast that inertial effects become relevant, invalidating our estimates for viscous friction. This comment applies to both filaments and membranes, but does not change the conclusion that the compression modes relax much faster than bending modes~\cite{LangerSeifert1993}.} Note that the bending relaxation time depends strongly on length scale; a ripple with a 1\,$\mu$m wavelength relaxes 1000 times faster than a ripple with a 10\,$\mu$m wavelength. Thus, the time scale for bending dynamics of lipid membranes can easily be of the order of seconds.

In addition to the viscous forces from the solvent, there are two other dissipative forces acting on the membrane: viscous forces due the surface viscosity of the membrane itself, and friction forces arising from the slipping of the two monolayers past each other. The relative importance of these forces changes with scale, with bilayer friction and then membrane viscosity becoming dominant as the length scale decreases~\cite{LangerSeifert1993,seifert1997}.  The implications of bilayer structure, both for dynamics and for morphology, are described in~\cite{seifert1997} and will not be discussed in this review. However, we will treat intrinsic membrane viscosity, since it is even more important than solvent viscosity in the case of polymersomes. 

\begin{figure}
\centerline{%\centerbmp{4.95 in}{2.84 in}{hyperact-fig.bmp}
\includegraphics[height=2.0in]{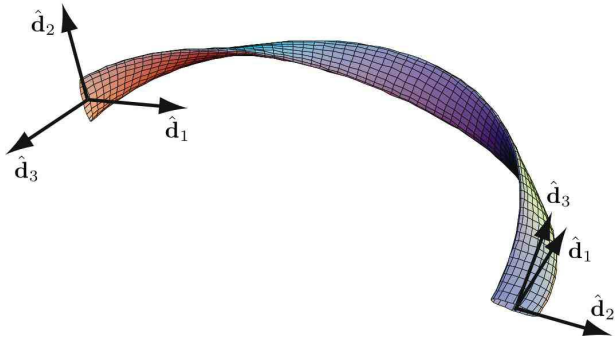}}
\caption{(Color online.) The directors for a rod are chosen along the principal axes of the cross-section of the rod.}
\label{rod_directors}
\end{figure}

Surface viscosity relates the viscous force per unit  \textit{length} $f_\mathrm{v}$ to the shear rate, $f_\mathrm{v}\sim\mu\nabla v$. Thus, the ratio of surface and bulk viscosity is a length, $\ell_\mathrm{SD}=\mu/\eta$, termed the ``Saffman-Delbr\"uck length"~\cite{Henle_etal2008} in honor of Saffman and Delbr\"uck's study of the mobility of particles embedded in a thin liquid film atop a deep liquid subphase~\cite{SaffmanDelbruck1975}. Since the force per unit area acting on an element of membrane is given by a derivative of the force per length, the surface viscous force per unit area scales as $\mu q^2 v$ compared to $\eta q v$ for the traction from the solvent. Therefore, surface viscosity effects become dominant when $\ell_\mathrm{SD}q>1$. Surface viscosities for lipid membranes are of the order of 10$^{-9}$\,N$\cdot$s/m~\cite{Dimova_etal1999,DimovaPoulignyDietrich2000,Dimova_etal2006}, whereas the surface viscosity of diblock copolymer vesicles can be about 500 times higher~\cite{Dimova_etal2002}. Therefore, the Saffman-Delbr\"uck length is $\ell_\mathrm{SD}\approx 1$\,$\mu$m for liposomes, and $\ell_\mathrm{SD}\approx1$\,mm for polymersomes. Therefore, surface viscosity effects dominate the dynamics of polymersomes. However, we will see in \S\ref{sec:DynamicalEquationsMembrane} that solvent viscosity cannot be neglected in the dynamics of almost flat membranes. Nevertheless, the bending relaxation time for a polymersome may be obtained by the same argument that led to Eq.~(\ref{bendrelaxtime}) with $\mu q$ replacing $\eta$, leading to 
\begin{equation}
\tau_\mathrm{b}\sim\frac{\mu}{\kappa q^2},
\end{equation}
or $\tau_\mathrm{b}\sim\mu L^2/\kappa$ for a vesicle of size $L$. Since surface viscous stresses will resist the relaxation of both compression and bending modes, the relation $\tau_\mathrm{s}\ll\tau_\mathrm{b}$ still holds for polymersomes.

\section{FILAMENTS}
\label{sec:curves}
In this section we derive the equations of motion for flexible filaments in a viscous liquid~\cite{doi_edwards1986}. We begin with kinematics and review how to compactly describe bending and twisting using a body-fixed frame of directors. Then we use variational arguments to derive the elastic forces per unit length acting on an element of a rod in an arbitrary configuration. After a brief discussion of compatibility relations for rod strain and angular velocity, we conclude the section with an overview of the viscous forces acting on slender rods in the overdamped regime of Stokes flow.

\subsection{Rod kinematics }\label{sec:RodKinematics} The centerline of a rod is given by $\br(\xi)$, the position of the point with coordinate $\xi$. Since $\xi$ is fixed for a given material point, we call $\xi$ a material coordinate. Alternatively, $\xi$ is sometimes called a convected coordinate since the label $\xi$ is carried along with the material point as the body deforms.  As mentioned earlier, it is much easier to bend a rod than it is to stretch it. Thus, we consider inextensible rods.  Note that $\xi$ is a material or convected coordinate; it is a label for material points of the centerline of the rod.  It is convenient to take the parameter $\xi$ to run from $0$ to $1$ from one end of the rod to the other. The distance $\mathrm{d}s$ between two nearby points on the rod centerline is given by the Pythagorean theorem and the chain rule,
\begin{equation}
\mathrm{d}s^2=\dd\br^2=\frac{\partial\br}{\partial \xi}\cdot\frac{\partial\br}{\partial \xi}\mathrm{d}\xi^2.\label{ds1d}
\end{equation}
The vector $\partial_\xi\br=\partial\br/\partial\xi$ is tangent to the centerline; normalizing this vector leads to the unit tangent vector
$\br'\equiv\partial_s\br$. 

Our characterization of the configuration of a rod is completed by specifying the orientation of the rod cross-section at each $s$. Let $\bd_1$ and $\bd_2$ lie along the principal directions of the rod cross-section (see Fig.~\ref{rod_directors}). Then define $\bd_3=\bd_1\times\bd_2$. For simplicity, we will consider \textit{unshearable} rods, for which the cross-section always remains perpendicular to the rod centerline, or $\bd_3=\br'$~\cite{antman1995}. Note that if the cross-section is circular, then we must arbitrarily choose a direction $\bd_1$ at each point $s$. For the case of a rod which is straight in the absence of external forces, a sensible choice is to make $\bd_1(s)$ constant.

\begin{figure}
\centerline{%\centerbmp{4.95 in}{2.84 in}{hyperact-fig.bmp}
\includegraphics[height=2.5in]{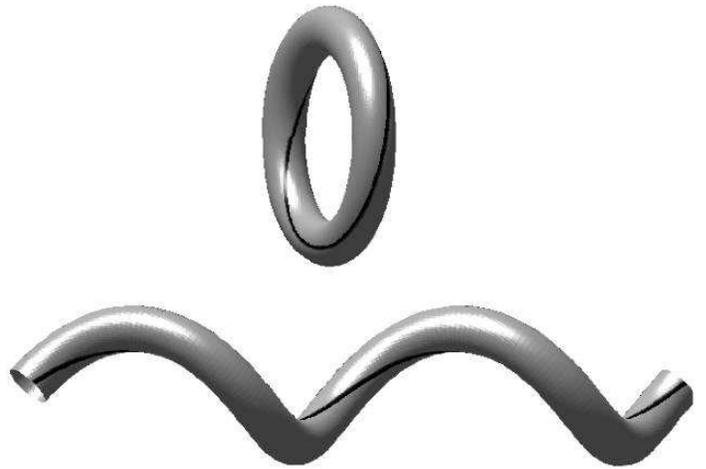}}
\caption{Two rods with the same  curvature $\kappa$ and the same  twist $\Omega_3$. For each rod, the black line traces the path of the tip of $\bd_1$. Note that the fact that one rod is closed and one is open is immaterial for determining the twist. }
\label{circlehelix}
\end{figure}

Let the subscripts $a$, $b$, ... $=1$, $2$, or $3$ label the members of orthonormal frames. The configuration of the rod is given by the orientation of the frame for all $s$. Instead of specifying the orientation of the frame, it is much more convenient to describe the configuration of the rod in terms of how fast the orthonormal frame rotates as $s$ increases.  The frames for any two nearby points differ by a small rotation, 
\begin{equation}
\bd_a'=\bOm\times\bd_a,\label{Omstrain}
\end{equation}
where in the body-fixed frame of directors, $\bOm=\Omega_1\da+\Omega_2\db+\Omega_3\dc$. We will use the Einstein convention of summing repeating indices to save space: $\bOm=\Omega_a\bd_a$. The infinitesimal rotations $\Omega_a$ are uniquely defined once the directors are chosen, since $\Omega_1=\hat\mathbf{d}_2'\cdot\hat\mathbf{d}_3$, and $\Omega_2$ and $\Omega_3$ are determined by cyclically permuting the indices.
The quantities $\Omega_1$ and $\Omega_2$ are the components of the curvature of the centerline, and $\Omega_3$ is the rate at which the cross-section twists around $\bd_3$ as $s$ increases.
For comparison, it is useful to recall the Serret-Frenet frame curve of a space curve, $\{\hat\mathbf{T}, \hat\mathbf{N}, \hat\mathbf{B}\}$, where $\hat\mathbf{T}=\br'$ is the unit tangent vector, and the normal $\hat\mathbf{N}$ and the binormal $\hat\mathbf{B}$ are defined by  
\begin{eqnarray}
\hat\mathbf{N}'&=&\tau\hat\mathbf{B}-\kappa\hat\mathbf{T}\label{dN}\\
\hat\mathbf{B}'&=&-\tau\hat\mathbf{N}\label{dB}\\
\hat\mathbf{T}'&=&\kappa\hat\mathbf{N}\label{dT}.
\end{eqnarray}
\begin{figure}
\centerline{%\centerbmp{4.95 in}{2.84 in}{hyperact-fig.bmp}
\includegraphics[height=1.6in]{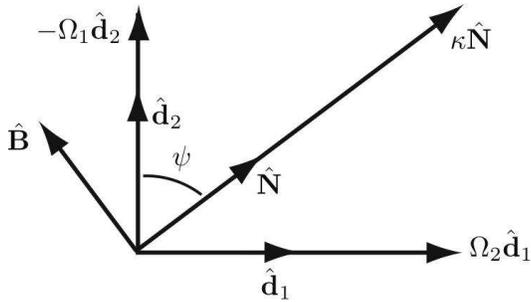}}
\caption{The basis vectors of the material and Serret-Frenet frames in the plane of the rod cross-section.}
\label{FSandMaterial}
\end{figure}
The curvature $\kappa$ is the rate of rotation of $\hat\mathbf{T}$ about $\hat\mathbf{B}$, and the torsion $\tau$ is the rate of rotation of $\hat\mathbf{N}$ about $\hat\mathbf{T}$. It is conventional to take $\kappa\ge0$ to make $\hat\mathbf{N}$ point to the center of curvature. Comparison of Eqs.~(\ref{Omstrain}) and (\ref{dN}--\ref{dT}) reveals that $\kappa^2=\Omega_1^2+\Omega_2^2$ and $\Omega_3=\psi'+\tau$, where $\psi$ is the angle between $\hat\mathbf{N}$ and $\bd_2$: $\bd_2\cdot\hat\mathbf{N}=\cos\psi$ (Fig.~\ref{FSandMaterial}). Note that the directions of the normal and binormal become ambiguous when the rod is straight, $\kappa=0$. This ambiguity can be sidestepped by using the so-called natural frame, which is well-defined even at points where $\kappa=0$ ~\cite{bishop1975,goldstein_etal1998,wolgemuth_et_al2004}. However, there is no ambiguity about $\bd_1$ and $\bd_2$ when the rod is straight; note further that $\Omega_1$ and $\Omega_2$ can have either sign. In this review, we consider rods which are straight in absence of external forces. Thus, the $\Omega_a$ may be interpreted as strains associated with bending and twisting. Note that a rod with $\bOm=\kappa_0\bd_2$ is an arc of a circle, and a rod with $\bOm=\kappa_0\bd_2+\tau_0\bd_3$ is a helix. Note also that all three variables $\Omega_a$ must be given to specify the shape of a rod; specifying only the curvature and twist does not determine the shape. For example, 
\begin{equation}
\bOm_{\mathrm{circle}}=\frac{1}{R}\cos(s/R)\bd_1+\frac{1}{R}\sin(s/R)\bd_2+\frac{1}{R}\bd_3
\end{equation}
describes a rod bent into a circle with curvature $1/R$ and twist $1/R$, whereas 
\begin{equation}
\bOm_{\mathrm{helix}}=\frac{1}{R}\bd_2+\frac{1}{R}\bd_3
\end{equation} 
describes a helix that also has curvature $1/R$ and twist $1/R$
 (see Fig.~\ref{circlehelix}).

\subsection{Energies, forces, and moments}

\begin{figure}
\centerline{%\centerbmp{4.95 in}{2.84 in}{hyperact-fig.bmp}
\includegraphics[height=0.9in]{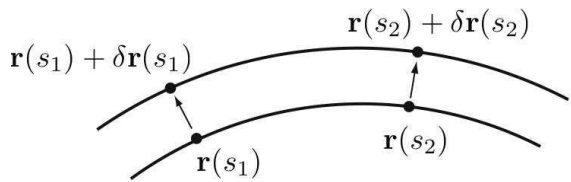}}
\caption{In general, the variation of a curve stretches the curve.}
\label{rdr}
\end{figure}

Our next task is to find the force $\beff\dd s$ exerted on an element of the rod of length $\dd s$ by the rest of the rod. This force equals the force the element exerts on the surrounding medium. It is useful to first consider a simple but classic question: what is the force per unit length that a curve in the plane exerts on its surroundings when its energy function is $E=\gamma\int\dd s$? Suppose also that the curve encloses a fixed area. Thus our problem models a two-dimensional droplet of incompressible fluid. To find the force for an arbitrary configuration $\br(s)$ of the curve, we impose an external force per unit length equal and opposite to $\beff(s)$ to put the instantaneous configuration in mechanical equilibrium. Thus, $\beff$ is given by the principle of virtual work, or the condition $\delta E_\mathrm{tot}=0$, where 
\begin{equation}
\delta E_\mathrm{tot}=\delta E+\int \beff(s)\cdot\delta\br(s)\dd s - p\delta A,\label{dE2d}
\end{equation}
and $p$ is the Lagrange multiplier enforcing the constraint of fixed enclosed area $A$. To calculate $\delta E_\mathrm{tot}$, we must find how the length $\dd s$ changes under a variation of the curve (Fig~\ref{rdr}): 
\begin{eqnarray}
\delta\dd s&=&\sqrt{[\br'+(\delta\br)']^2}\dd s-\dd s\nonumber\\
&=&\br'\cdot(\delta\br)'\dd s,\label{deltads}
\end{eqnarray}
where the second line of (\ref{deltads}) is accurate to first order in $\delta\br$. Note that $\br'=\dc$ is the unit tangent vector. Our formula correctly shows that the length element is invariant under rigid motions, such as the translations $\delta\br=\mathbf{a}$ and the rotations $\delta\br=\btheta\times\br$, where $\mathbf{a}$ and $\btheta$ are constant vectors. Using $\delta\dd s$ to calculate $\delta E_\mathrm{tot}$ yields
\begin{eqnarray}
\delta E_\mathrm{tot}&=&\gamma\int\br'\cdot(\delta\br)'\dd s+\int\left(\beff - p\hat\mathbf{N}\right)\cdot\delta\br\dd s\nonumber\\
&=&\int\left(-\gamma\br''+\beff-p\hat\mathbf{N}\right)\cdot\delta\br\dd s,
\end{eqnarray}
where we integrated by parts and used the fact that $\delta\br(L)=\delta\br(0)$ for a closed curve. Since $\delta\br$ is an arbitrary variation, the force is given by 
\begin{equation}
\beff=(\gamma \kappa + p)\hat\mathbf{N}.
\end{equation}
Note that the curve can only support normal forces per unit length. In equilibrium, $\beff=0$, and we recover the Young-Laplace law, with the Lagrange multiplier $p$ identified as the difference in the pressure outside the curve and inside the curve. Since $p$ is constant, $\kappa$ is constant, and the length-minimizing curve is a circle.

We now apply this approach to inextensible rods that resist bending~\cite{HallatschekFreyKroy2007} and twist. Unlike the previous example of a one-dimensional interface in the plane, an elastic rod has internal structure with its state defined by the strains $\Omega_a$. Suppose the elastic energy of the rod is given by 
\begin{equation}
E=\int {\cal E}(\Omega_1,\Omega_2,\Omega_3)\dd s.\label{genE}
\end{equation}
For example, the energy density 
\begin{equation}
{\cal E}=\frac{A}{2}\left(\Omega_1^2+\Omega_2^2\right)+\frac{C}{2}\Omega_3^2 =\frac{A}{2}\kappa^2+\frac{C}{2}\Omega_3^2\label{st_rodE}
\end{equation}
describes a rod with bend modulus $A$, twist modulus $C$, and a straight shape in the unstressed configuration. Our formalism is easily generalized to handle models in which ${\cal E}$ depends on derivatives of the $\Omega_a$~\cite{KanWolgemuth2007}, such as models for polymorphic phase transformations in bacterial flagella~\cite{goldstein_etal2000,CoombsHuberKesslerGoldstein2002,SrigirirajuPowers2005,SrigirirajuPowers2005b}. However, we will disregard such ``strain-gradient" terms for simplicity.

It is natural to suppose that the ends of the rod are subject to external forces $\bF_\mathrm{ext}(0)$ and $\bF_\mathrm{ext}(L)$,  and moments $\bM_\mathrm{ext}(0)$ and $\bM_\mathrm{ext}(L)$. To hold an arbitrary configuration in equilibrium with these boundary conditions, we must apply an external force per unit length $-\beff(s)$. Since the rod can twist about the tangent $\bd_3$, we must also apply an external moment per unit length $-m(s)\bd_3$. As in the closed curve example, $\beff(s)$ and $m(s)\bd_3$ are the forces and torques per unit length that the rod exerts on the medium for a given configuration.  These are again given by the condition $\delta E_\mathrm{tot}=0$, where $\delta E_\mathrm{tot}$ includes the work done by all external forces. Note that we are not considering the most general form of the external torque per unit length. A torque per unit length parallel to the centerline of the rod arises naturally when considering rotation of a filament in a viscous liquid (see \S\ref{illustrative1}), but our formalism can be easily modified to handle external torques per unit length that are perpendicular to the centerline. 

Since the rod is inextensible, care must be taken when performing the variation, since most variations $\delta\br(s)$ stretch the rod (Fig.~\ref{rdr}). The three components of $\delta\br$ are not independent for a variation that conserves the distance between material points on the rod. It is convenient to introduce a Lagrange multiplier function enforcing $\br'\cdot\br'=1$ and then vary the energy with respect to the three components of $\delta\br$ independently~\cite{GoldsteinLanger1995}. Instead we will follow the less common procedure of dispensing with the Lagrange multiplier function and constraining the variation~\cite{HallatschekFreyKroy2007}.  Since the constrained variation does not stretch the rod, arclength is conserved,
\begin{equation}
\delta\dd s=0.\label{dsiszero}
\end{equation}
Therefore, the order of applying the variation $\delta$ and derivatives with respect to $s$ is immaterial:
\begin{equation}
\delta \partial_s = \partial_s\delta.\label{deltads_commute}
\end{equation}
We may drop the distinction between $\delta(\br')$ and $(\delta\br)'$, and simply write $\delta\br'$.
 
\begin{figure}
\centerline{%\centerbmp{4.95 in}{2.84 in}{hyperact-fig.bmp}
\includegraphics[height=2.4in]{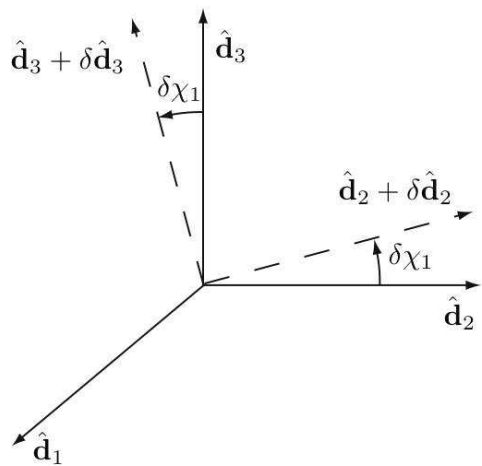}}
\caption{Rotation of directors under a variation with $\delta\chi_1\neq0$, $\delta\chi_2=\delta\chi_3=0$.}
\label{dchifig}
\end{figure}

Since the energy depends on the $\Omega_a$ only, the configuration of the rod is completely determined by the orientation of the directors in space, $\bd_a$.  Thus, we may vary the configuration without stretching the rod by subjecting the orthonormal frame of directors at each point $s$ to an infinitesimal rotation, $\delta\bd_a=
\beps\times\bd_a$ where $\beps\equiv\delta\chi_a\bd_a$. The infinitesimal angles $\delta\chi_a$ are related to the directors via 
\begin{eqnarray}
\delta\chi_1=\delta\bd_2\cdot\bd_3,\label{dchi1}
\end{eqnarray}
with $\delta\chi_2$ and $\delta\chi_3$ determined by cyclically permuting the indices in~(\ref{dchi1}). Thus, under a variation, the orthonormal frame vectors $\bd_2$ and $\bd_3$ rotate about $\bd_1$ through the angle $\delta\chi_1$ (Fig.~\ref{dchifig}). The two angles $\delta\chi_1$ and $\delta\chi_2$ correspond to the two degrees of freedom of $\delta\br$ under the constraint $\br'\cdot\br'=1$:
\begin{eqnarray}
\delta\chi_1=-\db\cdot(\delta\br)'\label{deltachi1}\\
\delta\chi_2=\da\cdot(\delta\br)'.\label{deltachi2}
\end{eqnarray}
In general, the are no globally defined angles $\chi_a$ whose variations correspond to the infinitesimal angles $\delta\chi_a$. The exceptions are cases when one of the directors is constant for the whole rod. For example, we may take $\bd_1$ to be normal to the plane containing a planar curve; then $\chi_1$ may be defined as the angle between the local tangent vector and a fixed direction in the plane, and $\chi_1'$ is the curvature. Likewise, the director $\bd_3$ has a constant direction for a straight, twisted rod, with $\chi_3$ the cumulative angle through which $\bd_1$ has rotated, and $\chi_3'$ the twist $\Omega_3$. 

The variation in the strain is found by varying Eq.~(\ref{Omstrain}), using Eq.~(\ref{deltads_commute}),  $\delta\bd_a=
\beps\times\bd_a$, and $\beps\equiv\delta\chi_a\bd_a$, and finally noting that $\mathbf{V}\times\bd_a=0$ for all $a$ implies $\mathbf{V}=0$:
\begin{equation}
\delta\bOm=\beps'+\beps\times\bOm.\label{vecdOm}
\end{equation}
It is also useful to write the variation in component form:
\begin{eqnarray}
\delta\Omega_1&=&\delta\chi_1'+\Omega_2\delta\chi_3-\Omega_3\delta\chi_2\label{dO1}\\
\delta\Omega_2&=&\delta\chi_2'+\Omega_3\delta\chi_1-\Omega_1\delta\chi_3\label{dO2}\\
\delta\Omega_3&=&\delta\chi_3'+\Omega_1\delta\chi_2-\Omega_2\delta\chi_1.\label{dO3}
\end{eqnarray}
The components of $\bOm\times\beps$ and not $\beps\times\bOm$ appear on the right-hand sides of Eqs.~(\ref{dO1}--\ref{dO3}), since the basis vectors $\bd_a$ in $\delta\bOm$ and $\beps$ must be varied and differentiated, respectively. 
To interpret these formulas, note that a variation with $\beps=\delta\chi_3(s)\bd_3$ changes the twist $\Omega_3$ and the principal curvatures $\Omega_1$ and $\Omega_2$, but not $\kappa=\sqrt{\Omega_1^2+\Omega_2^2}$. A deformation of the centerline, $\beps=\delta\chi_1(s)\bd_1+\delta\chi_2(s)\bd_2$ changes the twist if the rod is curved. This fact underlies the coupling of twist and writhe~\cite{caluguareanu1959,white1969,Fuller1971,kamien1998,goldstein_etal1998,DennisHannay2005}.

We now have all the necessary ingredients to find the force per unit length $\beff$. We  demand $\delta E_\mathrm{tot}=0$, where
\begin{eqnarray}
\delta E_\mathrm{tot}&=&\delta E+\int\left(\beff\cdot\delta\br+ m\delta\chi_3\right)\dd s\nonumber\\
&-&\bF_\mathrm{ext}(L)\cdot\delta\br(L)-\bF_\mathrm{ext}(0)\cdot\delta\br(0)\nonumber\\
&-&\bM_\mathrm{ext}(L)\cdot\beps(L)-\bM_\mathrm{ext}(0)\cdot\beps(0).\label{dEtotrod}
\end{eqnarray}
Since the energy is a function of $\Omega_a$ only, we have 
\begin{equation}
\delta E=\int M_a\delta\Omega_a \dd s,
\end{equation}
where $M_a\equiv\partial{\cal E}/\partial\Omega_a$. The formulas for the variation of the strain, Eqs.~(\ref{dO1}--\ref{dO3}), imply that 
\begin{eqnarray}
M_a\delta\Omega_a&=&M_a\delta\chi_a'+\bM\cdot\bOm\times\beps\label{MdOm1}\\
&=&(M_a\delta\chi_a)'-M_a'\delta\chi_a-\beps\cdot\bOm\times\bM\label{MdOm2}\\
&=&(M_a\delta\chi_a)'-\bM'\cdot\beps.\label{MdOm3}
\end{eqnarray}
Likewise, if we introduce the force $\bF(s)$ acting on the cross-section of the rod at $s$  by the relation $\bF'=\beff$, we have
\begin{eqnarray}
\beff\cdot\delta\br&=&\bF'\cdot\delta\br\\
&=&(\bF\cdot\delta\br)'-\bF\cdot\delta\br'\\
&=&(\bF\cdot\delta\br)'-\beps\cdot\bd_3\times\bF.\label{fdr3}
\end{eqnarray}
Combining Eqs.~(\ref{MdOm3}), (\ref{fdr3}), and (\ref{dEtotrod}), and demanding $\delta E_\mathrm{tot}=0$ for arbitrary (inextensible) $\delta\br$ yields $\bF(L)=\bF_\mathrm{ext}(L)$, $\bF(0)=-\bF_\mathrm{ext}(0)$, $\bM(L)=\bM_\mathrm{ext}(L)$, $\bM(0)=-\bM_\mathrm{ext}(0)$, and 
\begin{equation}
m\bd_3=\bM'+\bd_3\times\bF.\label{mombal}
\end{equation}
The boundary conditions on $\bM$ lead us to identify $\bM(s)$ as the moment exerted through the cross section at $s$ by the part of the rod with arclength greater than $s$ on the part with arclength less than $s$. Note that when $m=0$, Eq.~(\ref{mombal}) is the equilibrium moment balance equation for a rod~\cite{landau_lifshitz_elas}.

For the rod energy density~(\ref{st_rodE}), the relation $M_a=\partial{\cal E}/\partial\Omega_a$ yields $\bM=A(\Omega_1\bd_1+\Omega_2\bd_2)+C\Omega_3\bd_3$, or
\begin{equation}
\bM=A\bd_3\times\bd_3'+C\Omega_3\bd_3.\label{Melas}
\end{equation}
The transverse part of Eq.~(\ref{mombal}) yields
\begin{equation}
\bF=-A\br'''+C\Omega_3\br'\times\br''+\Lambda\br'.\label{F}
\end{equation}
Our variational procedure does not determine the function $\Lambda$. The value of $\Lambda$ is chosen to enforce the constraint $\br'\cdot\br'=1$, but from Eq.~(\ref{F}) we can see that it may be interpreted as a tension [see~\textcite{Deutsch1988} or~\textcite{GoldsteinLanger1995}, for example]. Note that $-A\br'''$ has a tangential part.

To summarize this section, we have used the principle of virtual work to derive the moment $\bM$ and force $\bF$ acting through any given 
cross-section of a rod with energy density~(\ref{st_rodE}). The constitutive relation $M_a=\partial\mathcal{E}/\partial\Omega_a$ arises naturally from the variation. Likwise, the conditions of force balance $\bF'=\mathbf{0}$ and moment balance $\bM'+\bd_3\times\bF=\mathbf{0}$ in equilibrium, when $\beff=\mathbf{0}$ and $m=0$, are not imposed but are implied by the principle of virtual work. These statements can be generalized to arbitrary energy densities along a curve~\cite{StarostinvanderHeijden2007,StarostinvanderHeijden2009}.

\subsection{Compatibility relations for rods} 
The variational formula  
Eq.~(\ref{vecdOm})
leads to important relations between the angular velocity and strain. The local angular velocity vector $\bom$ is given by $\bom=\beps/\delta t$ as $\delta t\rightarrow0$, where the variation is interpreted as the change of the rod shape over a time interval $\delta t$. Thus $\delta\bd_a=\beps\times\bd_a$ implies
\begin{equation}
\frac{\partial \bd_a}{\partial t}=\bom\times\bd_a\label{ddt}.
\end{equation}
Likewise, since we may similarly define $\partial\bOm/\partial t=\delta\bOm/\delta t$ as $\delta t\rightarrow0$, Eq.~(\ref{vecdOm}) implies
\begin{equation}
\frac{\partial\bOm}{\partial t}=\bom'+\bom\times\bOm.\label{dOmisomprime}
\end{equation}
These compatibility relations are also conveniently derived using the rotation 
matrices that act on the director frames~\cite{wolgemuth_et_al2004}. A third way of deriving the compatibility relations is to recognize that since $s$ and $t$ are independent variables, the order of taking derivatives does not matter, and thus we must have $\partial^2\bd_a/\partial t\partial s=\partial^2\bd_a/\partial s\partial t$. We will see below that the 3-component of Eq.~(\ref{dOmisomprime}),
\begin{equation}
\frac{\partial \Omega_3}{\partial t}=\frac{\partial\omega_3}{\partial s}+\Omega_1\omega_2-\Omega_2\omega_1,\label{fortwist}
\end{equation}
is required to obtain the evolution equation for twist. 
Once again, it is useful to note that $\bom'$ contains derivatives of $\omega_a$ and derivatives of the directors when comparing Eq.~(\ref{fortwist}) and Eq.~(\ref{dOmisomprime}); compare with the discussion after Eqs.~(\ref{dO1}--\ref{dO3}).

We may use Eqs.~(\ref{Omstrain}) and (\ref{ddt}) to recast Eq.~(\ref{fortwist}) in terms of the shape of the filament $\br$:
\begin{equation}
\partial_t\Omega_3=\partial_s\omega_3+\br'\times\br''\cdot(\partial_t\br)'.\label{compatibility}
\end{equation}
This relation may be viewed as a conservation law for twist density $\Omega_3$, with  $\omega_3$ the twist current, and the second term on the right-hand side acting as a source for twist~\cite{KlapperTabor1994,kamien1998,goldstein_etal1998}. The first term may be understood by observing that the only way the twist in a straight rod can change is if the rate of rotation $\omega_3$ is not uniform along the rod. The second term reflects the fact that motions of the backbone that do not involve a rotational velocity $\omega_3$ can also change the twist. Such motions require a change in the writhe, where the writhe of a closed curve is computed by counting the number of self-crossing of the curve in a planar projection, and then averaging over all projections [see~\textcite{DennisHannay2005} and references therein]. Although writhe is defined for a closed curve, its variation has a local form and can be considered for an open curve~\cite{KlapperTabor1994,kamien1998,goldstein_etal1998}.

\subsection{Viscous forces: slender body theory and resistive force theory}
In this section we describe how to approximately calculate the forces acting on a slender body immersed in a viscous fluid, and give a slightly more quantitative discussion of viscous fluid dynamics than in \S\ref{sec:TimeFilament}. The reader is directed to \cite{LaugaPowers2009} and references therein for a more detailed treatment. The stress in an incompressible viscous fluid is related to the strain rate by $\sigma_{ij}=-p\delta_{ij}+\eta(\bnabla_i u_j+\bnabla_j u_i)$, where $p$ is the pressure, $\eta$ is the viscosity, and $\mathbf{u}$ is the flow velocity~\cite{landauFM}. Note that we use the indices $i$, $j$, ... for the Cartesian coordinates.
As argued in \S\ref{sec:TimeFilament}, viscous forces dominate at the small scales of polymer and cells.
Thus we are permitted to disregard inertia. In this limit, the flow is governed by the Stokes equations:
\begin{eqnarray}
\eta\nabla^2\mathbf{u}-\bnabla p&=&\mathbf{0}
\label{stokes1}\\
\bnabla\cdot\mathbf{u}&=&0.\label{stokes2}
\end{eqnarray}
Note that the pressure $p$ plays the same role as $\Lambda$ did in our discussion of inextensible polymers, Eq.~(\ref{F}): the value of $p$ is determined to enforce the constraint of incompressibility, $\bnabla\cdot\mathbf{u}=0$. The boundary conditions for the Stokes equations are that the velocity of the fluid at a solid boundary must match the velocity of the boundary.

Ideally, we would like to calculate the viscous forces acting on a moving and deforming filament. Since no time derivatives appear in the Stokes equations~(\ref{stokes1}--\ref{stokes2}), the flow $\mathbf{u}$ and thus the forces acting on the filament at a given instant depend only on the geometry of the filament---its velocity and shape at that instant. However, it is generally impossible to find analytic expressions for the forces for anything but the simplest shapes. The problem can be simplified by developing an approximation scheme that exploits the smallness of the ratio of the rod diameter to its length; this approach is known as slender-body theory~\cite{lighthill1976}. Slender-body theory relies on the linearity of the Stokes equations to construct solutions for the flow around a moving or deforming filament by superposition of singular solutions. These singular solutions are the flows induced by a point force or a dipole source, and are analogous to the solutions used in the method of images in electrostatics. The end result is that slender-body theory provides a method, typically numerical, for computing the hydrodynamic forces distributed along a filament. Representative examples include work by~\textcite{shelley_ueda2000} and~\textcite{Cortez2001}.

In the limit of an asymptotically thin filament, slender-body theory simplifies to a local theory known as resistive force theory~\cite{gray_hancock1955,chwang_wu1971,lighthill1976}. Resistive force theory is local in the sense that it describes the hydrodynamic force per unit length acting at a point on a filament in terms of the velocity of the filament at that point, in contrast with slender-body theory, which accounts for the fact that the motion of every point along the filament induces a flow that affects the hydrodynamic forces at any given point on the filament. Thus, it is often stated that resistive force theory does not properly account for hydrodynamic interactions. Nevertheless, since resistive force theory is much simpler to implement that slender-body theory, and because it can give accurate results, we will use resistive force theory to formulate the equations of motion of filaments.

In resistive force theory, the fluid is treated as a passive background material that does not respond to the motion of the slender object. 
The viscous force per unit length acting on the rod is anisotropic, and given by \begin{equation}
{\bf f}_\mathrm{v}=-\zeta_\perp[{\bf v}- (\bd_3\cdot{\bf v})\bd_3]-\zeta_{||}
(\bd_3\cdot{\bf v})\bd_3, \label{sbhf}
\end{equation}
where $\mathbf{v}$ is the velocity of the rod relative to the local velocity of the fluid. In the limit of small $a/\lambda$, the friction coefficients in Eq.~(\ref{sbhf}) are defined by 
\begin{eqnarray}
\zeta_\perp&\sim&4\pi\mu/\ln(\lambda/
a)\label{zperpasymp}\\
\zeta_{||}&\sim&2\pi\mu/\ln(\lambda/a)\label{zparasymp},
\end{eqnarray}
where  $a$ is the filament diameter, and $\lambda$ is a cutoff representing either the rod length or the characteristic radius of curvature of the rod~\cite{keller_rubinow1976}.  Note that replacing $\lambda$ by
$2\lambda$ in Eqs.~(\ref{zperpasymp}--\ref{zparasymp}) merely
leads to subdominant additive terms in the denominator.  The expression (\ref{sbhf}) becomes accurate
 when
$a$ becomes much smaller than the rod's length or
typical radius of curvature $\lambda$~\cite{keller_rubinow1976}. Macroscopic scale experiments  show that resistive force theory does surprisingly well in describing the shape of driven elastic filaments in highly viscous fluid~\cite{koehler_powers2000,YuLaugaHosoi2006,QianPowersBreuer2008}. Numerical calculations show that resistive force theory is inaccurate for tightly coiled helices in flow, where hydrodynamic interactions are expected to be more important~\cite{kim_powers2005}. Even in a quiescent fluid, resistive force theory can give qualitatively wrong predictions for the sense of rotation of a towed superhelix~\cite{JungMareckFauciShelley2007}. Thus, resistive force theory is usually justified in a first approach to a new problem, but its predictions must always be viewed with a critical eye.

In some situations, such as the rotation of a nearly straight rod about its axis~\cite{wolgemuth_et_al2000a}, it is necessary to include the
viscous moment per unit length for rotation about the local tangent vector $\bd_3$. The appropriate local approximation for this moment is the viscous torque per unit length on a rotating cylinder:
\begin{equation}
{\bf m}_\mathrm{v}=-\zeta_\mathrm{r}(\bd_3\cdot{\bm\omega})\bd_3,\label{sbhm}
\end{equation}
where $\zeta_r=4\pi\eta a^2$ is the rotational drag coefficient~\cite{landauFM}.
In other situations, such as the rotation of helix, the contribution to the torque due to the rigid body motion of the centerline dominates the contributions from Eq~(\ref{sbhm}).

\subsection{Equations of motion for curves} We are ready to combine all the parts we have introduced and arrive at the coupled equations for the evolution of a rod. Since we work in the highly overdamped regime, all forces must sum to zero:
\begin{equation} 
\bF'+\mathbf{f}_\mathrm{v} = \mathbf{0}.
\end{equation}
Suppose that there is no imposed flow in the fluid; thus the velocity of the rod relative to the fluid is $\mathbf{v}=\partial_t\br$. Using the elastic force on a cross section~(\ref{F}) and the viscous force per unit length~(\ref{sbhf}), we find
\begin{equation}
\zeta_\perp\left(\partial_t\br\right)_\perp+\zeta_{||}\left(\partial_t\br\right)_{||}=-A\br''''+C\left(\Omega_3\br'\times\br''\right)'+\left(\Lambda\br'\right)',\label{reqn}
\end{equation}
where $(\partial_t\br)_\perp=\partial_t\br -[\bd_3\cdot(\partial_t\br)]\bd_3$ and $(\partial_t\br)_{||}=[\bd_3\cdot(\partial_t\br)]\bd_3$. To determine $\Lambda$, write $\mathbf{v}$ in terms of the mobilities $\zeta_\perp$ and $\zeta_{||}$ and the internal force per unit length $\mathbf{f}=\mathbf{F}^\prime$,
\begin{equation}
\mathbf{v}=\frac{\mathbf{f}_{||}}{\zeta_{||}}+\frac{\mathbf{f}_\perp}{\zeta_\perp},
\end{equation}
and use $\partial_t(\mathbf{r}^\prime\cdot\mathbf{r}^\prime)=2\partial_t(\mathbf{r}^\prime)\cdot\mathbf{r}^\prime=2\partial_s(\mathbf{v})\cdot\mathbf{r}^\prime$ to find
\begin{equation}
\partial_s\beff\cdot\dc+\left(1-\frac{\zeta_{||}}{\zeta_\perp}\right)\beff\cdot\partial_s\dc=0.\label{LambdaEqn}
\end{equation}
Using Eq.~(\ref{F}) to calculate $\beff=\bf{F}^\prime$ yields the desired differential equation for $\Lambda$.

The elastic and viscous moments must also sum to zero:
\begin{equation}
\mathbf{M}'+\bd_3\times\bF+\mathbf{m}_\mathrm{v}=\mathbf{0}.\label{momentbalance}
\end{equation}
The only nontrivial terms in Eq.~(\ref{momentbalance}) are the tangential components, which simplify to 
\begin{equation}
C\Omega_3'=\zeta_\mathrm{r}\omega_3.\label{tanmbal}
\end{equation}
To find the evolution equation for twist, combine Eqs.~(\ref{tanmbal}) and (\ref{compatibility}) to find
\begin{equation}
\zeta_\mathrm{r}\partial_t\Omega_3=C\Omega_3''+\zeta_\mathrm{r}\br'\times\br''\cdot\left(\partial_t\br\right)',\label{twisteqn}
\end{equation}
where $\partial_t\br$ must be deduced from Eq.~(\ref{reqn}).

We close this section by recalling the characteristic  time scales for the motion of filaments in a viscous fluid described in \S\ref{sec:TimeFilament}. Examination of Eq.~(\ref{reqn}) for a nearly straight filament with no twist reveals that bending modes have a characteristic relaxation time $\tau_\mathrm{b}\sim\zeta_\perp L^4/A$. Likewise, Eq.~(\ref{twisteqn}) applied to a straight but twisted rod implies that twist deformations have a characteristic  relaxation time $\tau_\mathrm{t}=\zeta_\mathrm{r}L^2/C$. 

\subsection{Illustrative example: twirling and whirling of elastic filaments}\label{illustrative1}

\begin{figure}
\centerline{%\centerbmp{4.95 in}{2.84 in}{hyperact-fig.bmp}
\includegraphics[height=1.4in]{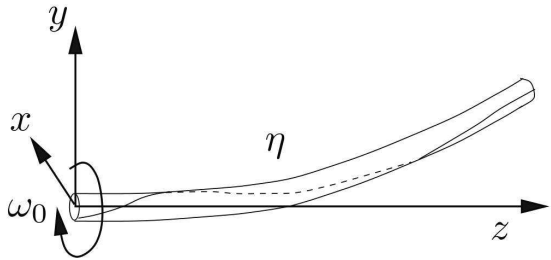}}
\caption{An elastic filament immersed in a liquid with viscosity $\eta$, and twirled about the $z$-axis with angular speed $\omega_0$. The straight state is unstable above a critical angular speed.}
\label{RotateFilamentFig}
\end{figure}

To illustrate the use of the equations of motion, we consider a fundamental problem: the whirling instability of a rotating elastic rod in a viscous fluid~\cite{wolgemuth_et_al2000a,LimPeskin2004,wada06}. 
A motor at $s=0$ rotates the rod about its long axis with angular speed $\omega_0$ (Fig.~\ref{RotateFilamentFig}). Initially, the rod is straight. Viscous forces cause twist to build up. For sufficiently high rotation speed, the straight state is unstable, and the rod writhes. Study of this problem will provide us with intuition that will help us understand more complicated phenomena in rotating filaments, such as the propagation of polymorphic instabilities in bacterial flagella~\cite{hotani1982,CoombsHuberKesslerGoldstein2002} or the chiral dynamics  of beating cilia in embryonic development~\cite{Nonaka_etal1998,HilfingerJulicher2008}.

As mentioned earlier, in this review we disregard the effects of thermal fluctuations.
We consider a rod with length much smaller than the bend persistence length and twist persistence length. We also suppose the length is much smaller than the twist persistence length $k_\mathrm{B}T/C$, which is defined by an argument similar to the one we gave in \S\ref{ThermalFlucts}. See \cite{wada06} for a study of the whirling instability for rods so long that thermal effects are important. 

Since we assume the rod is thin, it is natural to 
 use the local drag formulas Eq.~(\ref{sbhf}) and Eq.~(\ref{sbhm}) to describe the viscous forces and moments acting on the rod. Dropping the subscript 3 to reduce clutter, the balance of torques about the tangential direction (\ref{tanmbal}) becomes $C\Omega'=\zeta_\mathrm{r}\omega$. In the base state, the rod is straight with $\br=z\hat{\bf z}$, the tension $
 \Lambda$ vanishes, and the rod undergoes rigid-body rotation about the $z$-axis with speed $\omega=\omega_0$, leading to a constant gradient in twist. At the free end, $s=L$, the torque vanishes: $C\Omega=0$. Thus, 
\begin{equation}
\Omega=\zeta_\mathrm{r}\omega_0(s-L)/C.\label{ExTwistProfile}
\end{equation}
To estimate when the distributed viscous drag causes the rod to twist so much that the straight state becomes unstable, we may consider the related problem of a rod twisted by external moments acting at either end, in the absence of viscous forces. In this simpler problem, the straight state of the rod becomes unstable when the twist moment is roughly equal to the characteristic bending moment of a rod of length $L$: $C\Omega_\mathrm{crit}\sim A/L$~\cite{landau_lifshitz_elas}. Since $C\approx A$ for ordinary materials~\cite{landau_lifshitz_elas}, the rod writhes into a nonplanar shape once it is twisted by about one turn. In our problem with distributed external moments, the speed at which the typical twist torque balances the characteristic bending torque is given by 
$\zeta_\mathrm{r}\omega_\mathrm{crit}L\sim A/L$, or 
\begin{equation}
\omega_\mathrm{crit}\sim A/(\zeta_\mathrm{r}L^2).\label{ExOmegaCritEst}
\end{equation}
Since $C\approx A$, the critical rotation rate is the same order of magnitude as the relaxation rate for twist.

To find the precise value of the critical speed, linearize the equation of motion about the base state, Eq.~(\ref{reqn}):
\begin{equation}
\zeta_\perp\partial_t\mathbf{r}_\perp=-A\mathbf{r}^{\prime\prime\prime\prime}_{\perp}+C\hat\mathbf{z}\times\left(\Omega\mathbf{r}_\perp''\right)',\label{ExLinrEqn}
\end{equation}
where $\br_\perp=(x,y,0)$ and the twist is given by Eq.~(\ref{ExTwistProfile}). To leading order, the deflection of the rod is perpendicular to the $z$-axis. We therefore expect that the tension $\Lambda$ is second order in $x$ and $y$, which may be confirmed by expanding Eq.~(\ref{LambdaEqn}) in powers of $x$ and $y$.

Perturbations to the twist are governed by the diffusion equation with diffusion constant $C/\zeta_\mathrm{r}$. Thus, the perturbations in shape and twist may be studied independently. Since perturbations to twist are not destabilizing, we focus on Eq.~(\ref{ExLinrEqn}), which governs the shape of the filament. As in the static twist instability, this equation is most conveniently analyzed using complex notation $\xi=x+\mathrm{i}y$~\cite{landau_lifshitz_elas}. In these variables, Eq.~(\ref{ExLinrEqn}) becomes
\begin{equation}
\zeta_\perp\partial_t\xi=-A\xi^{\prime\prime\prime\prime}+\mathrm{i}\zeta_\mathrm{r}\omega_0\left[(s-L)\xi^{\prime\prime}\right]^\prime.\label{ExComplexEqn}
\end{equation}
For boundary conditions, we assume the end at $z=0$ is held fixed, $\xi(0,t)=0$, and clamped, $\xi'(0,t)=0$. At the far end the conditions are  vanishing  force, $\xi^{\prime\prime\prime}(L,t)=0$, and vanishing bending moment $\xi^{\prime\prime}(L,t)=0$. The solutions to Eq.~(\ref{ExComplexEqn}) are proportional to $\exp(\mathrm{i}\sigma t)$. When the imaginary part of $\sigma$ is negative, the base state is unstable; the real part of $\sigma$ is the rate of rigid body rotation of the filament about the $z$ axis. Equation~(\ref{ExComplexEqn}) has nontrivial solutions only for special values of $\omega_0$; to determine the value for which the straight state first becomes unstable, we assume $\xi(s,t)=\exp{\mathrm{i}\chi t}$, where $\chi$ is real, and use the boundary conditions to determine the critical $\omega_0$. This procedure is best done numerically. \textcite{wolgemuth_et_al2000a} found
\begin{equation}
\omega_\mathrm{crit}\approx8.9 \frac{A}{\zeta_\mathrm{r} L^2},\label{ExOmegaCritExact}
\end{equation}
in agreement with our rough estimate Eq.~(\ref{ExOmegaCritEst}). 
The motion of the filament can be described in terms of two different rotations. One is the rapid local rotation of each element of the rod about the local tangent vector; this rotation has rate $\omega_0$. But the centerline of the filament undergoes rigid-body motion at a different, much smaller rate $\chi$, which at
$\omega=\omega_\mathrm{crit}$ takes the value\footnote{\textcite{Wolg2009private} has provided a  more accurate value than the one stated in~\cite{wolgemuth_et_al2000a}.}
\begin{equation}
\chi\approx22.9\frac{A}{\zeta_\perp L^4}.
\end{equation}
The rigid-body rotation rate $\chi$ is of the same order of magnitude as the relaxation time for bending. 
The ratio of $\omega_\mathrm{crit}$ to $\chi$ is of order $L^2/a^2$ since the ratio of the corresponding resistance coefficients $\zeta_\perp/\zeta_\mathrm{r}$ is of order $L^2/a^2$.

Numerical methods applied to the full equations of motion are required to find the whirling state for $\omega_0>\omega_\mathrm{crit}$. \textcite{wolgemuth_et_al2000a} attempted a weakly nonlinear analysis by applying a numerical pseudospectral analysis~\cite{GoldsteinMurakiPetrich1996} to the equations of motion truncated at second order in displacement, and found a small-amplitude steady-state whirling state for $\omega$ just above $\omega_\mathrm{crit}$. However, subsequent work using the immersed boundary method~\cite{LimPeskin2004}, as well as simulations that include the full coupling of thermal, elastic, and hydrodynamic forces~\cite{wada06}, have shown that there is no small-amplitude whirling state, and that the weakly nonlinear analysis yields incorrect results. \textcite{wada06} found the same value for the critical frequency that we quote here; \textcite{Wolg2009private} has also found this value as well as the large amplitude whirling state using the pseudospectral methods employed in~\cite{wolgemuth_et_al2000a}. \textcite{LimPeskin2004} found a much lower value for the critical frequency. The reasons for the discrepancy remain unclear.

\section{MEMBRANES}
\label{sec:surfaces}
In this section we follow the same format as our discussion of curve dynamics, but we consider fluid membranes~\cite{boal2002}. We begin with the geometry of a fixed surface in space. This material is standard, but we include it for completeness. Then we turn to the calculation of membrane forces per unit area as a function of membrane shape. Since traditional approaches to this calculation are somewhat tedious, we review a recent idea of Guven that embodies economy of effort~\cite{Guven2004}. We then revisit the basic geometric concepts for a moving surface, taking care to distinguish between time dependence arising from flow or surface deformation and time dependence arising from time-dependent coordinates. Finally we conclude with the equations of motion for an incompressible fluid membrane and apply them to the instability of a cylindrical membrane tube under pressure.

\subsection{Geometry of a stationary surface}\label{sec:GeomSurface}

The intrinsic geometry of a space curve is Euclidean: a one-dimensional being confined to a space curve cannot tell if the curve bends and twists in the ambient three-dimensional space. This fact is why we can always choose arclength coordinates for a curve. In contrast, curved surfaces are non-Euclidean. Anyone who has attempted to flatten an orange peel onto a tabletop knows that it is impossible to construct a Cartesian coordinate system on a sphere. Thus we are forced to use curvilinear coordinates when a surface has curvature, and the description of the geometry of surfaces is necessarily more complicated than for curves. A key theme is that geometrical (and physical) concepts do not depend on the choice of coordinate system. Thus, we will gradually be led to the basic ideas of tensor calculus, which will allow us to construct geometrical quantities such as curvature in terms of any curvilinear coordinate system.

An additional complication arises for fluid surfaces that was not present for solid rods: even for a stationary shape, there may be a complex flow pattern on the surface. We will return to this complication later; first we will consider stationary surfaces and review the geometry necessary to calculate the force per unit area due to internal stresses as a function of membrane shape.

Denote the coordinates of a stationary surface as $\{\xi^1,\xi^2\}$.  Greek letters $\alpha$, $\beta$,  ... refer to the two-dimensional coordinates, such as $\xi^\alpha$, and Latin letters $i$, $j$, ...  continue to refer to Cartesian three-dimensional coordinates. Thus the position of a point in space on the surface is given by $\mathbf{r}(\xi^1,\xi^2)$. The distance between two points on the surface is given by the Pythagorean theorem  
\begin{equation}
\mathrm{d}s^2=\dd\br^2=\frac{\partial\mathbf{r}}{\partial\xi^\alpha}\cdot\frac{\partial\mathbf{r}}{\partial\xi^\beta}\mathrm{d}\xi^\alpha\mathrm{d}\xi^\beta.\label{fff}
\end{equation}
The right-hand side of (\ref{fff}) is known as the first fundamental form.  The quantity \begin{equation}
g_{\alpha\beta}=\partial_\alpha\br\cdot\partial_\beta\br
\end{equation}
is known as the metric tensor, because it converts distances $\{\dd\xi^1,\dd\xi^2\}$ measured in coordinate space to distance measured in real, three-dimensional space. 
Since distances can be measured by a being on a two-dimensional surface, the metric tensor is an intrinsic quantity.

A surface has two independent tangent vectors at any point, which may be taken to be  $\bt_\alpha\equiv\partial_\alpha\br$ along the direction of
increasing $\xi^\alpha$, for $\alpha=1,2$. 
In general,  the basis $\{\bt_1,\bt_2\}$ is not an orthonormal basis. The tangent vectors $\bt_\alpha$ are not geometric objects, since they depend on the choice of coordinates, but a tangent vector field $\mathbf{W}=W^\alpha\bt_\alpha$ is a geometric object independent of the choice of coordinates.
The unit normal to the surface is given by the cross product of the two coordinate tangent vectors, suitably normalized:
\begin{equation}
\nhat=\frac{\bt_1\times\bt_2}{|\bt_1\times\bt_2|}.
\end{equation}
Note that there are two unit normal vectors at every point, $\nhat$ and $-\nhat$. Unless there is an asymmetry such as different fluids on either side of the membrane, there is no reason to prefer $\nhat$ over $-\nhat$. 
The element of area on a surface is given by the parallelogram formed by the infinitesimal vectors $\bt_1\dd\xi^1$ and $\bt_2\dd\xi^2$, which has area
\begin{equation}
\dd S=\nhat\cdot\left(\bt_1\dd\xi^1\times\bt_2\dd\xi^2\right)=\sqrt{g}\dd\xi^1\dd\xi^2,
\end{equation}
where $g=g_{11}g_{22}-g_{12}g_{21}$ is the determinant of the metric tensor.

The curvature of a space curve is the tangential component of the rate of change of the normal to the curve [Eq.~(\ref{dN})]. Since a surface has two tangential directions, curvature of a surface is described by a bilinear form known as the second fundamental form:
\begin{equation}
\mathrm{d}\nhat\cdot\mathrm{d}\mathbf{r}=-K_{\alpha\beta}\mathrm{d}\xi^\alpha\mathrm{d}\xi^\beta,
\end{equation}
where
\begin{equation}
K_{\alpha\beta}=-\partial_\alpha\nhat\cdot\bt_\beta.\label{Kab}
\end{equation}
Note that since $\nhat\cdot\bt_\beta=0$, we have
\begin{equation}
\partial_\alpha\nhat\cdot\bt_\beta=-\nhat\cdot\partial_\alpha\bt_\beta=-\nhat\cdot\partial_\beta\bt_\alpha=\partial_\beta\nhat\cdot\bt_\alpha,\label{ndotdt}\end{equation}
and therefore conclude that the second fundamental form is symmetric: $K_{\alpha\beta}=K_{\beta\alpha}$. 

To make a stronger connection between our discussion of curves and the second fundamental form, consider a curve on a surface, parameterized by arclength, $\mathbf{x}(s)=\mathbf{r}(\xi^1(s),\xi^2(s))$, and define the normal curvature $\kappa_\mathrm{n}$ by $\kappa_\mathrm{n}=\nhat\cdot\kappa\hat\mathbf{N}=\nhat\cdot\mathrm{d}\hat\mathbf{T}/\mathrm{d}s$, or 
\begin{eqnarray}
\kappa_\mathrm{n}&=&\nhat\cdot\frac{\mathrm{d}\xi^\alpha}{\mathrm{d}s}\frac{\partial}{\partial\xi^\alpha}\left(\frac{\mathrm{d}\xi^\beta}{\mathrm{d}s}\bt_\beta\right)\\
&=&K_{\alpha\beta}\frac{\mathrm{d}\xi^\alpha}{\mathrm{d}s}\frac{\mathrm{d}\xi^\beta}{\mathrm{d}s}.
\end{eqnarray}
The normal curvature is the ratio of the second and the first fundamental forms~\cite{struik1988},
\begin{equation}
\kappa_\mathrm{n}=\frac{K_{\alpha\beta}\mathrm{d}\xi^\alpha\mathrm{d}\xi^\beta}{g_{\mu\nu}\mathrm{d}\xi^\mu\mathrm{d}\xi^\nu}.\label{normalcurvature}
\end{equation}
Equation~(\ref{normalcurvature}) implies that at a given point, all  curves  with the same direction $\mathrm{d}\xi^2/\mathrm{d}\xi^1$ and parameterized by arclength have the same normal curvature. 

It is useful to characterize the curvature of a surface by the maximum and minimum values of the normal curvature at a point. If $\mathbf{v}=v^\alpha\bt_\alpha$ is a unit vector, then the extreme values of $\kappa_\mathrm{n}$ are given by the critical points of
\begin{equation}
f=v^\alpha K_{\alpha\beta}v^\beta-\lambda (v^\alpha g_{\alpha\beta}v^\beta-1),
\end{equation}
where $\lambda$ is a Lagrange multiplier. Demanding that $\partial f/\partial v^\alpha=0$ yields
\begin{equation}
K_{\alpha\beta}v^\beta-\lambda g_{\alpha\beta}v^\beta=0.\label{extremevalues}
\end{equation}
Note that multiplying Eq.~(\ref{extremevalues}) by $v^\alpha$ and summing over $\alpha$, together with the definition of normal curvature (\ref{normalcurvature}), implies that 
$\lambda=\kappa_\mathrm{n}$. Since $v^\alpha\neq0$, the condition for Eq.~(\ref{extremevalues}) to have a solution is 
$\det(K_{\alpha\beta}-\kappa_\mathrm{n}g_{\alpha\beta})=0$. If we use $g^{\alpha\beta}$ to represent the inverse of the metric tensor, 
\begin{equation}
g^{\alpha\gamma}g_{\gamma\beta}=\delta^\alpha_{\ \beta},
\end{equation}
then what we have shown is that the maximum and minimum values of the normal curvature are given by the eigenvalues of the matrix
\begin{equation}
K^\alpha_{\ \beta}\equiv g^{\alpha\gamma}K_{\gamma\beta}.
\end{equation}
These extreme values are known as the principal curvatures of the surface. Not that since $K_{\alpha\beta}$ and $g_{\alpha\beta}$ are symmetric, the principal directions $v^\alpha$ corresponding to the two principal curvatures are orthogonal. Note that Eq.~(\ref{Kab}) may be written in terms of $K^\alpha_{\ \beta}$ to yield the Weingarten equations:
\begin{equation}
\partial_\alpha\nhat=-K_\alpha^{\ \beta}\that_\beta. \label{Weingarten}
\end{equation}

As an aside, we emphasize the importance of whether or not an index is raised or lowered. The convention is to use the metric tensor to lower indices, and the inverse of the metric tensor to raise indices. Quantities with upper and lower indices transform oppositely from one another under coordinate changes; for example, $W^\alpha$ and $\bt_\alpha$ transform under under coordinate changes in such as way that the vector field 
$\mathbf{W}=W^\alpha\bt_\alpha$ remains invariant. These transformation rules are reviewed in Appendix~\ref{coordtmns}.

Returning now to the discussion of curvature, we note that 
it is common to describe the curvature of a surface in terms of the
mean curvature, $H=K^\alpha_{\ \alpha}/2$, and the Gaussian curvature, $K=\det K^\alpha_{\ \beta}$. The mean curvature $H$ is the average of the sum of the two principal curvatures of a surface. It is a measure of how a surface bends in three-dimensional space; $H$ is analogous to the curvature $\kappa$ of a space curve. A plane has vanishing mean curvature, a cylinder has mean curvature equal to half the inverse of its radius, and a sphere has mean curvature equal to the inverse of its radius. The Gauss curvature $K$ is the product of the two principal curvatures of a surface. It is an \textit{intrinsic} quantity; $K$ can be determined by two-dimensional beings making local measurements of length on the surface. For example, by choosing appropriate coordinates, one can show that the relation between the circumference $C$ and radius $R$ of a small circle on a surface involves $K$ only~\cite{doCarmo1976}: $C\approx 2\pi R(1-KR^2/6)$ (Fig.~\ref{twoDcurvature}). Below we use direct calculation to see that $K$ is an intrinsic quantity.
Note that a plane and a cylinder both have vanishing Gauss curvature, which is consistent with the fact that local measurements confined to the surface cannot distinguish a plane from a cylinder. 

\begin{figure}
\centerline{%\centerbmp{4.95 in}{2.84 in}{hyperact-fig.bmp}
\includegraphics[height=2.3in]{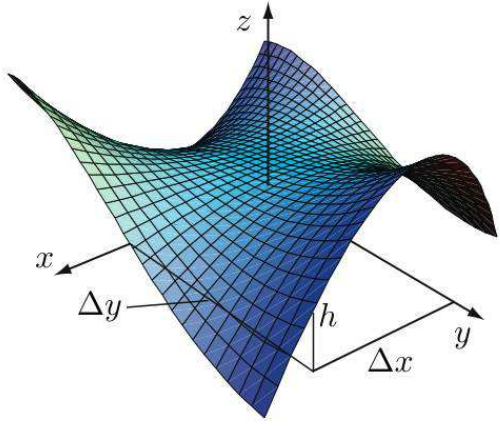}}
\caption{(Color online.) The height near the origin is given by the second fundamental form: $h\approx K_{\alpha\beta}\Delta\xi^\alpha\Delta\xi^\beta$, where $\Delta\xi^1=\Delta x$ and $\Delta\xi^2=\Delta y$.}
\label{hisKxy}
\end{figure}

In our study of curves we found it useful to express the derivative of directors in terms of the directors. Likewise, it is useful to examine the derivative of the coordinate tangent vector fields $\bt_\alpha$ in terms of $\bt_\alpha$ and $\nhat$.
Since the coordinate tangent vectors on a surface need not have unit length, the derivative of a tangent vector has both tangential and normal components: 
\begin{equation}
\partial_\alpha\bt_\beta=K_{\alpha\beta}\nhat+\Gamma^\gamma_{\alpha\beta}\bt_\gamma,\label{dadtb}
\end{equation}
where the normal components have been identified using the Weingarten equations~(\ref{Weingarten}) and $\nhat\cdot\bt_\alpha=0$, and the as yet unknown quantities $\Gamma^\gamma_{\alpha\beta}$ are called the Christoffel symbols. Note that $\Gamma^\gamma_{\alpha\beta}=\Gamma^\gamma_{\beta\alpha}$ since $\partial_\alpha\partial_\beta\mathbf{r}=\partial_\beta\partial_\alpha\mathbf{r}$ and  $K_{\alpha\beta}=K_{\beta\alpha}$.
The formula $K_{\alpha\beta}=\nhat\cdot\partial_\alpha\partial_\beta\mathbf{r}$ also suggests another interpretation of the second fundamental form. Choose a point 
$p$  on the surface and introduce Cartesian coordinates $\{x,y,z\}$ in three-dimensional space such that $z$ is parallel to $\nhat$. Near $p$, we may choose coordinates $\{\xi^1,\xi^2\}\approx\{x,y\}$. The height of the surface over the tangent plane is given by the second fundamental form (Fig.~\ref{hisKxy}). 

To see how the Christoffel symbols are related to the shape, observe that 
\begin{equation}
\partial_\alpha g_{\beta\gamma}=\partial_\alpha(\bt_\beta\cdot\bt_\gamma)=\Gamma_{\alpha\beta}^\delta g_{\delta\gamma}+\Gamma_{\alpha\gamma}^\delta g_{\delta\beta}.
\label{Ggab}\end{equation}
By cyclically permuting $\alpha$, $\beta$, and $\gamma$ in Eq.~(\ref{Ggab}) and then adding and subtracting appropriately, we can solve for the Christoffel symbols to find
\begin{equation}
\Gamma^\gamma_{\alpha\beta}=g^{\gamma\delta}\Gamma_{\delta\alpha\beta}=\frac{g^{\gamma\delta}}{2}\left(\partial_\alpha g_{\gamma\beta}+\partial_\beta g_{\gamma\alpha} - \partial_\gamma g_{\alpha\beta}\right).\label{Christoffel}
\end{equation}
Since the Christoffel symbols depend only on the metric and its derivatives, they are intrinsically quantities. 

The equations~(\ref{dadtb}) are sometimes called the Gauss equations or the partial differential equations of surfaces~\cite{struik1988}. These equations are indispensable for formulating the equations of motion for fluids on a deforming surface. In particular, we will need to take derivatives of the velocity field, which has tangential and normal components. However, we will limit this section to a simplified discussion in which we consider the derivative of a purely tangential vector field, $\mathbf{W}=W^\alpha\bt_\alpha$:
\begin{equation}
\partial_\alpha\mathbf{W}=\left(\partial_\alpha W^\beta\right)\bt_\beta+W^\beta\Gamma^\gamma_{\alpha\beta}\bt_\gamma+W^\beta K_{\alpha\beta}\nhat.
\end{equation}
Once the direction $\bt_\alpha$ of the derivative is chosen, the tangential and normal components of $\partial_\alpha\mathbf{W}$ are uniquely defined, independent of choice of coordinates. These quantities contain all the information about how $\mathbf{W}$ changes along the direction $\bt_\alpha$. In contrast, the quantities $\partial_\alpha W^\beta$ do not contain complete information about the change in $\mathbf{W}$. For example, we can imagine a vector field on small region of a curved surface in which the components $W^\beta$ are constants. Then $\partial_\alpha W^\beta=0$, but $\mathbf{W}$ is not constant, since the surface is curved.
Therefore, we are lead to define the \textit{covariant derivative} of the vector field $W^\beta$ as the tangential components of $\partial_\alpha\mathbf{W}$:
\begin{equation}
\left(\nabla_\alpha W\right)^\beta=\partial_\alpha W^\beta+\Gamma^\beta_{\alpha\gamma}W^\gamma.
\end{equation}
The covariant derivative of a tangent vector field is an intrinsic quantity since the Christoffel symbols are intrinsic. 

The covariant derivative is readily generalized to other quantities defined on a surface. For scalar functions, $\nabla_\alpha f=\partial_\alpha f$. Since each Cartesian component of a vector field may be considered a scalar function, we may also write $\nabla_\alpha \mathbf{W}=\partial_\alpha\mathbf{W}$. The Leibnitz rule is used to define the covariant derivative of higher-order tensors. For example, we may write the dot product of two vector fields $\mathbf{U}$ and $\mathbf{W}$ as $\mathbf{U}\cdot\mathbf{W}=U_\alpha W^\alpha$, where $U_\alpha=g_{\alpha\beta}U^\beta$. The Leibnitz rule then implies
\begin{equation}
\nabla_\alpha(U_\beta W^\beta)=(\nabla_\alpha U)_\beta W^\beta + U_\beta(\nabla_\alpha W)^\beta.\label{leibnitzUW}
\end{equation}
Since $U_\beta W^\beta$ is a scalar function,  $\nabla_\alpha(U_\beta W^\beta)=\partial_\alpha(U_\beta W^\beta)$ and solving (\ref{leibnitzUW}) for $(\nabla_\alpha W)_\beta$ yields
\begin{equation}
\left(\nabla_\alpha W\right)_\beta = \partial_\alpha W_\beta - \Gamma^\gamma_{\alpha\beta}W_\gamma.
\end{equation}
Since the Cartesian coordinates of a vector are scalar functions, we may write
\begin{equation}
\nabla_\alpha\bt_\beta=\partial_\alpha\bt_\beta-\Gamma^\gamma_{\alpha\beta}\bt_\gamma=K_{\alpha\beta}\nhat.\label{Dt}\end{equation} At first sight this relation may seem confusing since we defined the covariant derivative of $W^\beta$ by removing the normal part from $\partial_\alpha\mathbf{W}$, and here we have normal components in the covariant derivative of $\bt_\beta$. The key point is that the two-dimensional indices determine the nature of the covariant derivative.  The rule (\ref{Dt}) ensures that the Leibnitz rule holds for the components and basis vectors of a vector: $\nabla_\alpha\mathbf{W}=\partial_\alpha\mathbf{W}=(\nabla_\alpha W)^\beta\bt_\beta+W^\beta\nabla_\alpha\bt_\beta$.

The general rule is that the covariant derivative of a tensor has a term with a Christoffel symbol added for every upper index, and a term with a Christoffel symbol subtracted for every lower index. For example, 
\begin{equation}
\nabla_\alpha K^\beta_{\ \gamma}=\partial_\alpha K^\beta_{\ \gamma}+\Gamma^\beta_{\alpha\delta}K^\delta_{\ \gamma}-\Gamma^\delta_{\alpha\gamma}K^\beta_{\ \delta}.\end{equation}
Since the metric tensor $g_{\alpha\beta}$ is derived from the Euclidean metric of the ambient three-dimensional space, the covariant derivative of the metric tensor vanishes:
\begin{equation}
\nabla_\alpha g_{\beta\gamma}=\partial_\alpha g_{\beta\gamma}-\Gamma^\delta_{\alpha\beta}g_{\delta\gamma}-\Gamma^\delta_{\alpha\gamma}g_{\beta\delta}=0.\label{DgIs0}
\end{equation}
This fact may be deduced by writing Eq.~(\ref{leibnitzUW}) in terms of $U^\alpha$, $W^\beta$, and $g_{\alpha\beta}$. The metric tensor and its inverse is free to pass through the covariant derivative.

To formulate the generalization of the Stokes equations (\ref{stokes1}--\ref{stokes2}) for membrane flow on a curved surface we will need to consider multiple covariant derivatives acting on the velocity field. 
Special care must be taken, since covariant derivatives do not commute when acting on a vector field. 
The Gauss-Codazzi equations, which we now derive, determine the relation between $(\nabla_\alpha\nabla_\beta)W^\gamma$ and $(\nabla_\beta\nabla_\alpha)W^\gamma$. 
Since the tangent vector field $\mathbf{W}$ is three scalar functions on the surface,  covariant derivatives commute when acting on $\mathbf{W}$:
\begin{equation}
(\nabla_\alpha\nabla_\beta -\nabla_\beta\nabla_\alpha)\mathbf{W}=0.
\end{equation}
Resolving this formula into tangential and normal components yields the Gauss-Codazzi equations:
\begin{eqnarray}
(\nabla_\alpha\nabla_\beta-\nabla_\beta\nabla_\alpha)W^\gamma&=&
(K_{\ \alpha}^\gamma K_{\beta\delta}-K_{\ \beta}^\gamma K_{\alpha\delta})W^\delta\label{gauss}\\
\nabla_\alpha K_{\beta\gamma}&=&\nabla_\beta K_{\alpha\gamma}.\label{codazzi}
\end{eqnarray}
The right-hand side of Eq.~(\ref{gauss}) can be simplified by noting that $K_{\ \alpha}^\gamma K_{\beta}^{\ \delta}-K_{\ \beta}^\gamma K_{\alpha}^{\ \delta}$ is asymmetric in $(\alpha,\beta)$ and $(\gamma,\delta)$, and therefore proportional to the determinant of $K^\alpha_{\ \beta}$. The only independent component is $\alpha=1$, $\beta=2$, and we may write
\begin{eqnarray}
(\nabla_1\nabla_2-\nabla_2\nabla_1)W^1&=&K W_2\label{simpleGC}\\
(\nabla_1\nabla_2-\nabla_2\nabla_1)W^2&=&-K W_1.\label{simpleGC2}
\end{eqnarray}
The Gaussian curvature, originally defined in terms of the extrinsic quantity $K_{\alpha\beta}=\nhat\cdot\partial_\alpha\partial_\beta\br$, is intrinsic. The fact that $K$ may be determined by two-dimensional being living on a surface is known as Gauss's \textit{Theorema Egregium}~\cite{Morgan1998}.

We close this subsection with a discussion of the covariant form of Green's theorem for a two-dimensional surface. This relation will be required when we derive the stress tensor for a membrane. 
First note that  Eq.~(\ref{Christoffel}) implies that the covariant divergence of a two-dimensional vector field has a simple form: 
\begin{equation}
(\nabla_\alpha V)^\alpha = \frac{1}{\sqrt{g}}\partial_\alpha \left(\sqrt{g}V^\alpha\right).
\end{equation}
This formula is useful when integrating by parts, since the area element $\dd S=\sqrt{g}\dd^2\xi$ depends on $\xi^\alpha$:
\begin{equation}
\int V^\alpha \partial_\alpha f\dd S
=-\int(\nabla_\alpha V^\alpha)f\dd S+\int\nabla_\alpha(V^\alpha f)\dd S.\label{intbyparts}
\end{equation}
In appendix~\ref{green} we show that the second term of  Eq.~(\ref{intbyparts}) can be written as an integral over the boundary of the surface:
\begin{equation}
\int_D\nabla_\alpha B^\alpha\dd S=\oint_{\partial D} B^\alpha p_\alpha\dd s,\label{greentext}
\end{equation}
where $\hat\mathbf{p}=\partial\mathbf{r}/\partial s\times\nhat$ is the unit tangent vector perpendicular to the boundary $\partial D$.

\subsection{Energies and forces}

In this section we find the force exerted on a patch of a membrane by the rest of the membrane, or, equivalently, the force the patch exerts on the surrounding medium. We follow the geometric approach of \textcite{Guven2004}; see also 
~\cite{CapovillaGuven2002} and~\cite{CapovillaGuven2004}. For concreteness, we focus on fluid membranes; however, the approach may also be applied to membranes with internal degrees of freedom such as tilt order~\cite{MuellerDeserno2005}. The elastic energy for a fluid membrane takes the form
\begin{equation}
E=\int\left[\frac{\kappa}{2}(2H)^2+\bar\kappa K\right]\dd S,\label{helf-can}
\end{equation}
where $\kappa$ and $\bar\kappa$ are moduli with units of energy~\cite{canham1970,helfrich1973}. For a closed vesicle, the Gauss-Bonnet theorem implies that the integral of the Gaussian curvature $K$ over the surface is a topological invariant~\cite{kamien2002}, and therefore does not change under a variation of the shape. Henceforth we drop this term. 

To find the force $\beff\dd S$ on a patch for an arbitrary shape, we invoke the principle of virtual work as we did with filaments. We suppose the shape is held in equilibrium by an external force per area $-\beff(\xi^1,\xi^2)$ and examine the condition $\delta E_\mathrm{tot}=0$, where
\begin{equation}
\delta E_\mathrm{tot} = \delta E + \int\beff\cdot\delta\br\dd S.
\end{equation} 
Finding the variation of $E$ for surfaces is harder then it is for curves. Since  $E=\int{\cal E}(g_{\alpha\beta},K_{\alpha\beta})\dd S$, we must find $\delta g_{\alpha\beta}$ and $\delta K_{\alpha\beta}$ in terms of $\delta\br$. This direct approach is straightforward but tedious~\cite{zhong-can_helfrich1989}. \textcite{Guven2004} suggested introducing Lagrange multiplier functions which allow the variations $\br$, $g_{\alpha\beta}$, and $K_{\alpha\beta}$ to be taken independently:

\begin{eqnarray}
\delta W&=&\delta E+\int\beff\cdot\delta\br\dd S +\delta\big\{\int \bF^\alpha\cdot(\partial_\alpha\br-\bt_\alpha)\dd S\nonumber\\
&+&\int\left[\lambda^\alpha_\perp\bt_\alpha\cdot\nhat+\lambda_\mathrm{n}(\nhat^2-1)\right]\dd S\nonumber\\
&+&\int\left[\Lambda^{\alpha\beta}(K_{\alpha\beta}+\bt_\alpha\cdot\partial_\beta\nhat)\right.\nonumber\\
&&\left.{\left.-\ (T^{\alpha\beta}/2)(g_{\alpha\beta}-\bt_\alpha\cdot\bt_\beta)\right]}\dd S\right\}.\label{LagMults}\end{eqnarray}
The energy density ${\cal E}$ is now regarded not as function of $\br$ but instead as a function of $g_{\alpha\beta}$ and $K_{\alpha\beta}$. The condition of equilibrium is that $\delta W=0$ for the variations $\delta\br$, $\delta\bt_\alpha$, $\delta\nhat$, $\delta K_{\alpha\beta}$, and $\delta g_{\alpha\beta}$. Since the variables appear quadratically in the constraints, the variations are easy to compute. For example, the variation with respect to $\br$ implies that
\begin{equation}
\nabla_\alpha\bF^\alpha=\beff.
\end{equation}
Thus the Lagrange multiplier function $\bF^\alpha$ is the two dimensional stress tensor. To see why the stress is of mixed character, it is useful to apply Green's theorem (\ref{greentext}):
\begin{equation}
\int_D\nabla_\alpha \mathbf{F}^\alpha\dd S=\oint_{\partial D} \mathbf{F}^\alpha p_\alpha\dd s.
\end{equation}
The force per unit length acting through a contour in the surface is a three-dimensional vector, and the normal $\hat\mathbf{p}$ to the contour lies in the local tangent plane and is therefore a two-dimensional vector.

It is convenient to define the functional derivative $\delta W/\delta\mathbf{h}_\alpha$, where $\mathbf{h}_\alpha$ is a generic mixed tensor, as
\begin{equation}
\delta W=\int\frac{\delta W}{\delta\mathbf{h}_\alpha}\cdot\delta\mathbf{h}_\alpha\dd S.
\end{equation}
The condition $\delta W/\delta\mathbf{t}_\alpha=\mathbf{0}$ yields the stress in terms of Lagrange multipliers,
\begin{equation}
\bF^\alpha=\left(T^{\alpha\beta}-\Lambda^{\alpha\gamma}K_\gamma^{\ \beta}\right)\bt_\beta+\lambda_\perp^\alpha\nhat.
\end{equation}
Likewise,  the condition $\delta W/\delta\nhat=\mathbf{0}$ along with Eq.~(\ref{Dt}) gives $\lambda_\perp^\alpha$ and $\lambda_\mathrm{n}$ in terms of $\Lambda^{\alpha\beta}$:
\begin{eqnarray}
\lambda_\perp^\alpha&=&\nabla_\beta\Lambda^{\alpha\beta}\label{lambdaperp}\\
2\lambda_\mathrm{n}&=&K_{\alpha\beta}\Lambda^{\alpha\beta}.\label{lambdan}
\end{eqnarray}
Finally, $\delta W/\delta K_{\alpha\beta}=0$ implies
\begin{equation}
\Lambda^{\alpha\beta}=-{\cal E}^{\alpha\beta},
\end{equation}
where ${\cal E}^{\alpha\beta}=\partial{\cal E}/\partial K_{\alpha\beta}$.

We need a few identities to help us find the variation of $E$ with respect to the metric tensor. For example, to find the variation of the determinant of the metric tensor, we use the identity 
\begin{equation}
\exp\log\det g_{\alpha\beta}=\exp\mathrm{tr}\log g_{\alpha\beta},
\end{equation}
which yields
\begin{equation}
\delta g = g g^{\alpha\beta}\delta g_{\alpha\beta}.
\end{equation}
Thus $\delta \dd S=g^{\alpha\beta}\delta g_{\alpha\beta}\dd S/2$. Note also that Eq.~(\ref{g-ginv}) implies 
\begin{equation}
\delta g^{\alpha\beta}=-g^{\alpha\gamma}\delta g_{\gamma\delta}g^{\delta\beta}.\label{dgab}
\end{equation}
With these formulas we see that $\delta W/\delta g_{\alpha\beta}=0$ implies
\begin{equation}
T^{\alpha\beta}=g^{\alpha\beta}{\cal E}+2\frac{\partial{\cal E}}{\partial g_{\alpha\beta}}.\label{Tab}
\end{equation}
The two-dimensional stress tensor may also be written as
\begin{equation}
T^{\alpha\beta}=\frac{2}{\sqrt{g}}\frac{\partial}{\partial g_{\alpha\beta}}\left(\sqrt{g}{\cal E}\right).
\end{equation}
Note  that the contributions from varying $\dd S$ in the constraint terms vanish since the constraints are invoked after taking the variation. To summarize, we have found a general formula for the stress tensor of a membrane in terms of ${\cal E}^{\alpha\beta}=\partial{\cal E}/\partial K_{\alpha\beta}$, the curvature tensor $K_{\alpha\beta}$, and the two-dimensional stress $T^{\alpha\beta}$~\cite{Guven2004}: 
\begin{equation}
\bF^\alpha=(T^{\alpha\beta}+{\cal E}^{\alpha\gamma}K_\gamma^{\ \beta})\bt_\beta-(\nabla_\beta{\cal E}^{\alpha\beta})\nhat.
\end{equation}
The variational approach with auxiliary variables can also be used to find the moments per length acting across any infinitesimal line segment in the membrane~\cite{CapovillaGuven2002,MullerDesernoGuven2007}. 
Two examples illustrate the formalism.

\subsubsection{Soap film}
The energy of a soap film is given by the interfacial tension $\gamma$ times the area. Thus, the energy density  ${\cal E}=\gamma$, and ${\cal E}^{\alpha\beta}=0$. The stress depends only on the two-dimensional tensor $T^{\alpha\beta}=\gamma g^{\alpha\beta}$, and
\begin{equation}
\bF_\mathrm{film}^\alpha=\gamma g^{\alpha\beta}\bt_\beta.\label{soapstress}
\end{equation}
$\bF_\mathrm{film}^\alpha$ is in the tangent direction as expected.
The force per unit area acting on the film is $\beff_\mathrm{film}=\nabla_\alpha \bF^\alpha_\mathrm{film}$ or
\begin{equation}
\beff_\mathrm{film}=2\gamma H\nhat+(\partial_\alpha\gamma)g^{\alpha\beta}\bt_\beta.\label{fsoap}
\end{equation}
The tangential component of Eq.~(\ref{fsoap}) corresponds to a Marangoni stress and arises when surfactant concentration varies over the film.
For soap film on a frame in equilibrium, $\gamma$ is uniform and $\beff=0$ is the minimal surface equation $H=0$. If the film forms a closed surface with fixed enclosed volume, we introduce another Lagrange multiplier as in Eq.~(\ref{dE2d}) and find that the mean curvature must be constant.

\subsubsection{Fluid membrane}\label{sec:fluidmembrane}

Consider the fluid membrane energy  (\ref{helf-can}) with $\bar\kappa=0$. The energy density is
${\cal E}=(\kappa/2)(g^{\alpha\beta}K_{\alpha\beta})^2$. To find ${\cal E}^{\alpha\beta}$, we must differentiate ${\cal E}$  with respect to $g_{\alpha\beta}$ while holding $K_{\alpha\beta}$ fixed.  Note that  Eq.~(\ref{dgab}) implies
\begin{equation}
\frac{\partial g^{\alpha\beta}}{\partial g_{\gamma\delta}}=-\frac{1}{2}g^{\alpha\gamma}g^{\beta\delta}-\frac{1}{2}g^{\alpha\delta}g^{\beta\gamma},
\end{equation}
where we have used the symmetry $g_{\alpha\beta}=g_{\beta\alpha}$.
A short calculation reveals that the bending stress in a fluid membrane is
\begin{equation}
\bF_\mathrm{bend}^\alpha=2\kappa\left[\left(g^{\alpha\beta}H^2-HK^{\alpha\beta}\right)\bt_\beta-\left(\nabla^\alpha H\right)\nhat\right].\label{Fabend}
\end{equation}
The force per unit area acting on the membrane due to internal bending stresses is given by  the divergence of the stress tensor $\bF_\mathrm{bend}^\alpha$,
\begin{equation}
\beff_\mathrm{bend}=-2\kappa\left(\nabla^2H+2H^3-2HK\right)\nhat.\label{fmembrane}
\end{equation}
We used Eq.~(\ref{codazzi}) to show that the tangential components of $\beff_\mathrm{bend}$ vanish for constant $\kappa$.
There is no tangential component of the force per unit area because the bending energy is invariant under changes in coordinates, and a small change in coordinates corresponds to a deformation of the surface along tangent directions at every point, 
\begin{equation}
\br(\xi^\alpha+\epsilon^\alpha)\approx\br(\xi^\alpha)+\epsilon^\alpha\bt_\alpha.
\end{equation}
Equation~(\ref{fmembrane}) for $\beff_\mathrm{bend}$ may be derived directly from $\beff=-\delta E/\partial\br$ with considerably more effort~\cite{zhong-can_helfrich1989}. 

Although we argued that membranes in the tense regime are approximately incompressible, it is instructive to examine the stress and force per unit area arising from the stretching energy,
 $E_\mathrm{s}=\int\mathcal{E}_\mathrm{s}\dd S$, where $\mathcal{E}_\mathrm{s}=(k/2)(\phi/\phi_0-1)^2$. First we will work by analogy with a two-dimensional liquid and identify the 
two-dimensional pressure $p_{2\mathrm{d}}$ by considering a situation with uniform density $\phi$~\cite{CaiLubensky1995}. In that case, the two-dimensional pressure is $p_{2\mathrm{d}}=-\partial E_\mathrm{s}/\partial A$, where $A$ is the area and the derivative is taken at fixed number  $\phi A$.  
Carrying out the differentiation leads to $p_{2\mathrm{d}}=-(\mathcal{E}_\mathrm{s}-\phi\mathcal{E}_\mathrm{s}')$, where $\mathcal{E}_\mathrm{s}$ is regarded as a function of $\phi$.

To apply the principle of virtual work to find the stress and elastic force, we must keep the identity of the particles fixed under the variation. This constraint arises because the principle of virtual work is derived by assuming mechanical equilibrium of a fixed set of particles.  Therefore, the variation satisfies $\delta(\sqrt{g}\phi)=0$, which implies $\delta\phi=-(1/2)g^{\alpha\beta}\delta g_{\alpha\beta}\phi$ and
\begin{equation}
\delta E_\mathrm{s}=-\int\frac{1}{2}\sqrt{g}g^{\alpha\beta}\delta g_{\alpha\beta}p_{2\mathrm{d}}\dd S.
\end{equation}
From Eq.~(\ref{Tab}), the corresponding two-dimensional stress tensor is $T_\mathrm{s}^{\alpha\beta}=-p_{2\mathrm{d}} g^{\alpha\beta}$ and we see that the two-dimensional pressure acts like a negative interfacial tension [compare with Eq.~(\ref{soapstress})]. Thus, $\bF_\mathrm{s}^\alpha=-p_{2\mathrm{d}}g^{\alpha\beta}\bt_\beta$ and $\beff_\mathrm{s}=-2p_{2\mathrm{d}}H\nhat-(\partial_{\alpha}p_{2\mathrm{d}})g^{\alpha\beta}\bt$.

We can make a further simplification by observing that the large modulus $k$ implies that the density will be close to its preferred value: $\phi=\phi_0+\phi_1$, where $\phi_1\ll\phi_0$. Working to first order in $\phi_1$ leads to 
 $p_\mathrm{2d}=k\phi_1/\phi_0$.  The stress arising from stretching is therefore 
\begin{equation}
\mathbf{F}^\alpha_\mathrm{s}=- \left(k\phi_1/\phi_0\right)g^{\alpha\beta}\bt_\beta =-\gamma g^{\alpha\beta}\bt_\beta,\label{Fas}
\end{equation}
where $\gamma=-p_{2\mathrm{d}}$.
In general, the stress $\bF^\alpha_\mathrm{s}$ cannot be disregarded in a membrane, since the smallness of the stretch $\phi_1$ is offset by the largeness of the modulus $k$. 
Note that in mechanical equilibrium, the tangential stresses acting on the membrane are constant, and therefore $p_{2\mathrm{d}}$ and $\phi_1$ are uniform, independent of the shape of the membrane. Often, membrane shapes are studied under the assumption of incompressibility, $k\rightarrow\infty$, in which case $\gamma=-p_{2\mathrm{d}}$ is treated as a constant Lagrange multiplier, and $\nabla_\alpha\bF^\alpha_\mathrm{s}$ is the associated force per area area enforcing the constraint. For dynamics, the Lagrange multiplier may have spatial dependence if there are tangential stresses.

\begin{figure}
\centerline{%\centerbmp{4.95 in}{2.84 in}{hyperact-fig.bmp}
\includegraphics[height=1.15in]{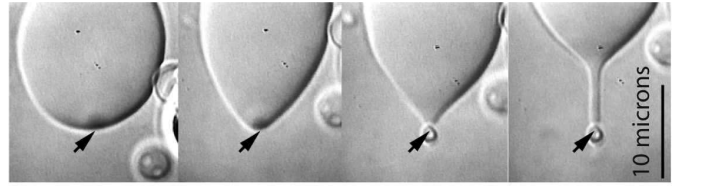}}
\caption{Equilibrium shapes of a vesicle subject deformed with an optical tweezer~\cite{fygenson_marko_libchaber1997}. The magnitude of the force increases from left to right. Figure courtesy of D. Fygenson.}
\label{tubule}
\end{figure}
The formulas for the stress tensors $\bF^\alpha_\mathrm{bend}$~(\ref{Fabend}) and $\bF^\alpha_\mathrm{s}$~(\ref{Fas}) give insight into the mechanics of membranes. Consider a tubule pulled out of a vesicle with a point force, (Fig.~\ref{tubule})~\cite{evans_yeung1994,fygenson_marko_libchaber1997}. It is common to regard the spherical part of the vesicle in the right-most frame of Fig.~\ref{tubule} as a reservoir of lipid, and to suppose there is an interfacial tension $\gamma$ that remains constant during the deformation. Equivalently, we may regard the membrane as under tension, with $\gamma=-p_{2\mathrm{d}}>0$ and constant. Since the spherical part of the vesicle is much bigger than the tubule, we suppose that $p_{2\mathrm{d}}$ is independent of tubule length once the tubule is formed. Thus, ${\cal E}=\gamma+(\kappa/2)(2H)^2$. The cylindrical part of the vesicle is easy to analyze, since $K=0$ and $H=-1/(2R)$, where $R$ is the tubule radius. We may disregard the pressure jump across the membrane since the large radius of the spherical part of the vesicle ensures that the pressure jump is small. The total force on an infinitesimal patch of the cylinder therefore vanishes, $\beff=\beff_\mathrm{film}+\beff_\mathrm{bend}=0$, or
\begin{equation}
-\gamma/R+\kappa/(2R^3)=0.
\end{equation}
In equilibrium, $R=\sqrt{\kappa/(2\gamma)}$.

\begin{figure}
\centerline{%\centerbmp{4.95 in}{2.84 in}{hyperact-fig.bmp}
\includegraphics[height=1.5in]{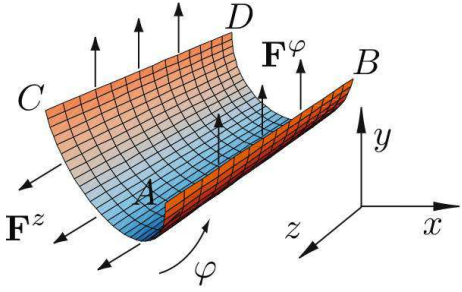}}
\caption{(Color online.) A section of a tubule, showing the forces per unit length $\bF^z$ and $\bF^\varphi$.}
\label{half-cyl}
\end{figure}

Closer inspection of the equilibrium configuration may seem to reveal a paradox~\cite{WaughHochmuth1987,powers_huber_goldstein2002,Fournier2007}. With the usual cylindrical coordinates $\{\xi^1,\xi^2\}=\{\varphi,z\}$, we expect that the interfacial tension $\gamma$ will lead to a force per unit length $\bF^\varphi=\gamma\bphihat$ along the edge $AB$ of Fig.~\ref{half-cyl}, with a force per unit length of equal magnitude along the other edge, $CD$. But since there is no pressure jump across the surface, there are no other forces to balance these forces. To resolve this apparent paradox, examine the stress $\bF^\alpha=\bF^\alpha_\mathrm{film}+\bF^\alpha_\mathrm{bend}$. In cylindrical coordinates, the first fundamental form is $\dd s^2=R^2\dd\varphi^2+\dd z^2$, and the second fundamental form is $K_{\alpha\beta}\dd\xi^\alpha\dd\xi^\beta=-R\dd\varphi^2$. Along a circular contour of constant $z$,
$\bF^z= 2\gamma\hat\mathbf{z}$. Thus, the force required to hold the tether in equilibrium is $4\pi\gamma\hat\mathbf{z}$; twice what it would be for a cylinder with no bending stiffness. However, along a line of longitude, constant $\varphi$, $\bF^\varphi=[\gamma-\kappa/(2 R^2)]\bphihat=0$; the contributions to the stress from the bending energy cancel the contributions from the interfacial energy, as they must since there is no pressure difference.

\subsection{Moving surfaces: kinematics}

\subsubsection{Coordinates on deforming surfaces with flow}
The description of a moving surface which may have internal flows requires careful thought. Curves are much simpler.  We saw that we can always use arclength for the spatial coordinate, and in the case of an inextensible filament, the arclength coordinate of a material point does not change. There is no analog of arclength parameterization for surfaces, and instead we must choose coordinates that are most appropriate for the situation. There are several common choices, including convective or material coordinates, surface-fixed coordinates, and general coordinates.  Material and surface-fixed coordinates are useful for formulating equations of motion~\cite{Aris1989}, while general coordinates are useful when one must solve for the evolution of a shape. In this section we briefly describe the most important aspects of each class of coordinates.

To define material or convective coordinates, parametrize the surface at time $t=0$ with coordinates $u^\Gamma=\{u^I,u^{II}\}$. We will use Roman numerals or upper-case Greek letters for material coordinates. Since we disregard thermal effects such as Brownian motion of the molecules making up the membrane, the motion of the material points defines a smooth flow velocity. The coordinates $u^\Gamma$ are labels that stay with the material particles as time evolves, with the trajectory of the material point $\{u^I,u^{II}\}$ given by  $\br(u^I,u^{II},t)$.  The velocity of a material point is given by 
\begin{equation}
\mathbf{V}=\left.\frac{\partial \mathbf{r}}{\partial t}\right|_{u^I,u^{II}}=V^\Gamma\bt_\Gamma+V_\mathrm{n}\nhat.
\end{equation}
As mentioned above, convected coordinates are useful for formulating the equations of motion. For example, we will see in \S\ref{sec:surfaces}D that the Reynolds transport theorem for conservation of particles is conveniently derived using convective coordinates~\cite{Aris1989}. However, since the motion of a surface is typically something to be solved for, $\br(u^I,u^{II},t)$ is usually unknown at the outset of any problem. 

An alternate approach is to attempt to define a coordinate system that is fixed to the surface, independent of the flows of the molecules making up the surface.  Such a choice is easy enough for a surface that does not change shape, but is impossible  for a deforming surface. The best we can do is to choose coordinates $\{\xi^1,\xi^2\}$  such that the velocity of a point with fixed coordinate---not necessarily a material point---is purely normal to the surface:
\begin{equation}
\left.\bt_\alpha\cdot\frac{\partial\br}{\partial t}\right|_{\xi^1,\xi^2}=0.\label{fixedframe}
\end{equation}
Thus, 
\begin{equation}
\mathbf{V}=\left.\frac{\dd \mathbf{r}}{\dd t}\right|_{u^I,u^{II}}=\bt_\alpha V^\alpha+V_\mathrm{n}\nhat,
\end{equation}
where $V^\alpha=\dd\xi^\alpha/\dd t$, $\xi^\alpha(t)$ is the trajectory of a particle with constant $u^\Gamma$, and $V_\mathrm{n}\nhat=\partial\mathbf{r}/\partial t$.  Note that $\partial\mathbf{r}/\partial t$ is purely normal. 
Surface-fixed coordinates are conceptually useful when formulating equations of motion of moving surfaces. For example, we will see below how the condition~(\ref{fixedframe}) leads to alternate expression the continuity equation that commonly appears in the literature~\cite{Aris1989,Youhei1994,YouheiErratum1997,DorriesFotin1996,Buzza2002}.  But like material coordinates, surface-fixed coordinates  also require knowledge of the shape evolution which is usually unknown until the complete problem is solved. 

Therefore, it is common to use general coordinates when one is solving for the shape of a membrane as a function of time.  By general coordinates we mean coordinates in the Eulerian point of view~\cite{ChaikinLubensky1995}, which are not constrained by conditions such as (\ref{fixedframe}) or the Lagrangian label of convected coordinates. For example, we will use cylindrical coordinates $\mathbf{r}(\theta,z,t)=h(z,t)\hat\mathbf{r}+z\hat\mathbf{z}$ in our example of axisymmetric surfaces. (Here $\hat\mathbf{r}$ is a unit vector in the radial direction in cylindrical coordinates.)  The velocity in general coordinates is
\begin{equation}
\mathbf{V}=\left.\frac{\dd \mathbf{r}}{\dd t}\right|_{u^I,u^{II}}=\frac{\partial\mathbf{r}}{\partial u^\alpha}\frac{\dd u^\alpha}{\dd t}+\frac{\partial\mathbf{r}}{\partial t},
\end{equation}
where $\partial\mathbf{r}/\partial t$ has \textit{both} normal and tangential components.

\subsubsection{Strain rate}
To formulate the hydrodynamics of fluid membranes, we need to measure the  strain rate, defined as rate that the distance between two nearby points on the surface changes with time. If $\dd\mathbf{r}$ is the vector connecting the two nearby points at a given time, then after a time interval $\dd t$, the two points are connected by $\dd\mathbf{r}'=\dd\mathbf{r}+(\dd\mathbf{V})\dd t$. The strain rate tensor $\mathcal{S}_{\alpha\beta}$ is then defined by 
\begin{equation}
\dd s'^2=\dd s^2+2\dd t\,\dd\xi^\alpha\dd\xi^\beta \mathcal{S}_{\alpha\beta},
\end{equation}
where $\dd s'^2=\dd\mathbf{r}'\cdot\dd\mathbf{r}'$ and $\dd s^2=\dd\mathbf{r}\cdot\dd\mathbf{r}$.
Writing $\dd s'^2$ in terms of $\dd \mathbf{r}$ and $\mathbf{V}$ leads to 
\begin{equation}
\mathcal{S}_{\alpha\beta}=\frac{1}{2}\left(\partial_\alpha\mathbf{V}\cdot\bt_\beta+\partial_\beta\mathbf{V}\cdot\bt_\alpha\right).\label{extrinsicSab}
\end{equation}
Note that $\mathcal{S}_{\alpha\beta}$ is unchanged by rotation with constant, uniform rate $\bom$, $\mathbf{V}=\bom\times\br$, as required by the definition of strain~\cite{DorriesFotin1996}. Likewise, the strain is invariant under  translations $\br\mapsto\br+\mathbf{a}$ and Galilean boosts $\mathbf{V}\mapsto\mathbf{V}+\mathbf{V}_0$, where $\mathbf{a}$ and $\mathbf{V}_0$ are constant and uniform~\cite{DorriesFotin1996}. 
Applying the Gauss-Weingarten equations (\ref{dadtb},\ref{Weingarten}) to Eq.~(\ref{extrinsicSab}) yields an alternate form,
\begin{equation}
\mathcal{S}_{\alpha\beta}=\frac{1}{2}\left(\nabla_\alpha V_\beta+\nabla_\beta V_\alpha-2K_{\alpha\beta}V_\mathrm{n}\right).\label{strainrate}
\end{equation}
The first two terms the right-hand side of Eq.~(\ref{strainrate}) are the covariant generalization of the strain rate of a two-dimensional material. The third term represents the strain that takes place for example when a sphere inflates uniformly---even though there are no velocity gradients, there is strain whenever there is curvature and a normal velocity $V_\mathrm{n}$. Note that gradients in $V_\mathrm{n}$ do not lead to strain at first order in $\dd t$. 

\subsubsection{Equation of continuity} It is useful to formulate  the covariant version of the conservation of particles, or, more generally, formulate the covariant version of the Reynolds transport theorem~\cite{Aris1989}. To this end, consider a patch of material points $S(t)$ on a fluid membrane, and a scalar quantity $\phi$, which for concreteness can be taken as the number per unit area. We wish to calculate the rate of change of the total number of  particles in the patch $S(t)$, assuming that the particles never leave the membrane to enter the solvent. To eliminate the time-dependence of the patch region, use convected coordinates:
\begin{equation}
\frac{\dd}{\dd t}\int_{S(t)}\phi\dd S=\frac{\dd}{\dd t}\int_{S_0}\phi(\xi^\alpha(u^\Gamma,t),t)\sqrt{G}\dd u^I\dd u^{II},
\end{equation}
where $S_0=S(t=0)$, and $G$ is the determinant of the metric in the material coordinates $\{u^I,u^{II}\}$. The total time derivative of $\phi$ is given by the chain rule,
\begin{equation}
\frac{\dd\phi}{\dd t}=\partial_t\phi+V^\alpha\partial_\alpha\phi,
\end{equation}
where $V^\alpha$ 
is the tangential component of the velocity $\mathbf{V}=V^\alpha\bt_\alpha+V_\mathrm{n}\nhat$. To evaluate the time derivative of $\sqrt{G}$, note that Eq.~(\ref{dgab}) implies $\dd\sqrt{G}/\dd t=\sqrt{G}(G^{\Gamma\Delta}\dd G_{\Gamma\Delta}/\dd t)/2$, and $\dd G_{\Gamma\Delta}/\dd t=\partial_\Gamma\mathbf{V}\cdot\bt_\Delta+\bt_\Gamma\cdot\partial_\Delta\mathbf{V}$. Reverting to general coordinates, we have
\begin{equation}
G^{\Gamma\Delta}\frac{\dd G_{\Gamma\Delta}}{\dd t}=g^{\alpha\beta}\left(\partial_\alpha\mathbf{V}\cdot\bt_\beta+\partial_\beta\mathbf{V}\cdot\bt_\alpha\right).\label{GdG}
\end{equation}
We may now use the Gauss-Weingarten equations~(\ref{dadtb},\ref{Weingarten}) to deduce
\begin{equation}
\frac{\dd}{\dd t}\int_{S(t)}\phi\dd S=\int_{S(t)}\left[\frac{\dd\phi}{\dd t}+\phi\nabla_\alpha V^\alpha-2\phi H V_\mathrm{n}\right]\dd S,
\end{equation}
or 
\begin{equation}
\frac{\dd\phi}{\dd t}+\phi\nabla_\alpha V^\alpha-2\phi H V_\mathrm{n}=0.\label{ContEqn}
\end{equation}
The continuity equation (\ref{ContEqn}) is sometimes expressed in terms of derivatives of the metric tensor using surface-fixed coordinates~\cite{Aris1989,Youhei1994,YouheiErratum1997,DorriesFotin1996,Buzza2002}, in which the condition~(\ref{fixedframe}) can be used to show 
\begin{equation}
\frac{1}{\sqrt{g}}\frac{\partial}{\partial t}\left(\sqrt{g}\phi\right)+\nabla_\alpha\left(\phi V^\alpha\right)=0.\label{ContEqnII}
\end{equation}
However, since Eq.~(\ref{ContEqnII}) only holds for surface-fixed coordinates, and not for example in the cylindrical coordinates we use to solve our illustrative problem in \S\ref{sec:polymersomes}, we will use Eq.(\ref{ContEqn}) in this review. 

For an incompressible membrane, $\phi$ is constant, and the continuity equation simplifies to 
\begin{equation}
\nabla_\alpha V^\alpha=2HV_\mathrm{n}.\label{VVn}
\end{equation}
In terms of the strain rate, incompressibility implies $\mathcal{S}^\alpha_{\ \alpha}=0$.

\subsection{Dynamical equations for an incompressible membrane}
\label{sec:DynamicalEquationsMembrane}
There are several complementary approaches to the dynamics of membranes from the points of view of fluid mechanics and interfaces~\cite{Scriven1960,waxman1984,KralchevskyErikssonLjunggren1994,ArroyoDeSimone2009}, statistical mechanics~\cite{CaiLubensky1994,CaiLubensky1995,MiaoLomholtKleis2002}, and the director approach analogous to our treatment of rods~\cite{HuZhangE2007}. Instead of presenting the most general models, we discuss the simple case of an incompressible fluid membrane. We disregard the bilayer nature of the membrane and therefore do not consider effects such as bilayer friction~\cite{evans_yeung1995,LangerSeifert1993}.

 The form of the two-dimensional viscous stress tensor is derived the same way as the stress for three-dimensional fluid mechanics in flat space~\cite{Aris1989}. Disregarding the possibility of viscoelastic effects~\cite{Rey2006}, the stress must be a symmetric tensor linear in the velocity gradients:
\begin{equation}
T^{\alpha\beta}_\mathrm{vis}=A g^{\alpha\beta}+B^{\alpha\beta\gamma\delta}\mathcal{S}_{\gamma\delta},
\end{equation}
with $A$ linearly proportional to $\mathcal{S}^{\alpha}_{\ \alpha}$ and $B$ independent of $\mathcal{S}$. Since $\mathcal{S}^\alpha_{\ \alpha}=0$ for an incompressible membrane, only $B^{\alpha\beta\gamma\delta}$ enters $T^{\alpha\beta}_\mathrm{vis}$. The symmetries of the stress and rate of strain tensors imply that $B^{\alpha\beta\gamma\delta}$ is symmetric under exchange of $\alpha$ and $\beta$, and also $\gamma$ and $\delta$. In principle, we could build $B^{\alpha\beta\gamma\delta}$ out of combinations of both of the tensors $K^{\alpha\beta}$ and $g^{\alpha\beta}$. However, we will work to leading order in an expansion in curvature, in which case
\begin{equation}
B^{\alpha\beta\gamma\delta}=\mu(g^{\alpha\gamma}g^{\beta\delta}+g^{\alpha\delta}g^{\beta\gamma})+\zeta g^{\alpha\beta}g^{\gamma\delta}.
\end{equation}
The constant $\zeta$ is the dilational viscosity, which does not enter the stress for an incompressible membrane since $\mathcal{S}^\alpha_\alpha=0$.
Thus,
$
T^{\alpha\beta}_\mathrm{vis}=2\mu\mathcal{S}^{\alpha\beta}
$, and the total membrane viscous stress is
\begin{equation}
\bF^\alpha_\mathrm{vis}=\mu\left(\nabla^\alpha V^\beta+\nabla^\beta V^\alpha-2K^{\alpha\beta}V_\mathrm{n}\right)\bt_\beta.\label{totmemvisstr}
\end{equation}
The force per unit area acting on the membrane due to the internal viscous stress is
$\beff_\mathrm{vis}=\nabla_\alpha\bF^\alpha_\mathrm{vis}$, or
\begin{eqnarray}
\beff_\mathrm{vis}&=&\mu[\nabla^2 V^\beta+KV^\beta-2V_\mathrm{n}\nabla^\beta H\nonumber\\
&+&2\left(Hg^{\alpha\beta}-K^{\alpha\beta}\right)\nabla_\alpha V_\mathrm{n}
]\bt_\beta\nonumber\\
&+&2\mu[K_{\alpha\beta}\nabla^\alpha V^\beta -V_\mathrm{n}(4H^2-2K)]\nhat,\label{fvisint}
\end{eqnarray}
where the Gauss-Codazzi equations~(\ref{simpleGC}) have been used to simplify the final form.
Various limits and forms of $\beff_\mathrm{vis}$ have appeared in the literature~\cite{Aris1989,Scriven1960,waxman1984,ArroyoDeSimone2009,DorriesFotin1996,Youhei1994,YouheiErratum1997}.
Note that Eq.~(\ref{fvisint}) leads to a qualitative change in the equations for fluid flow even in the simplest case of flow on a fixed sphere, for which $\beff_\mathrm{vis}=\mu(\nabla^2V^\beta+KV^\beta)\bt_\beta$, where $K$ is a constant~\cite{Henle_etal2008}.

It is natural to ask if there are any other terms of the same order in gradients of velocity that should be included in the surface viscous stress. For example, the term $\bF^\alpha_\mathrm{n}=\mu_2\nhat\nabla^\alpha V_\mathrm{n}$ with ``transverse" viscosity $\mu_2$ has often been used  in the theory of capillary waves on surfactant-laden interfaces~\cite{Goodrich1961II}.  However, this term has recently been shown to be unphysical by~\textcite{Buzza2002}. Buzza's argument is that $\bF^\alpha_\mathrm{n}$ does not obey  frame invariance. Frame invariance is a fundamental assumption of continuum mechanics that states that the relation between stress and strain for a small element of material cannot depend on the motion of the element~\cite{oldroyd1950,larson1988}, since at the scale of the microstructure, inertial effects are small for typical shear rates. In particular, the relation between stress and strain for a small element must remain the same if the element is subject to a time-dependent rotation. Removing the spurious term $\bF_\mathrm{n}^\alpha$ eliminates unphysical results such as negative measured dilatational viscosities that have plagued the interpretation of quasi-elastic light scattering experiments for many years~\cite{Buzza2002,CicutaHopkinson2004}.

To see why $\bF_\mathrm{n}^\alpha$ is unphysical, one can show that in a rotating frame, the force per unit volume depends on the rate of rotation for a rigid-body rotation~\cite{Buzza2002}. Perhaps a more direct argument that something is amiss with $\bF_\mathrm{n}^\alpha$ is that it predicts a positive dissipation rate for rigid-body rotation of a flat membrane~\cite{Buzza2002}.  If we include the $\mu_1$ term, then the dissipation rate
\begin{equation}
\dot W=\int\nabla_\alpha\mathbf{V}\cdot\bF_\mathrm{vis}^\alpha\dd S,
\end{equation}
has a term in the integrand proportional to $\mu_2(\nabla_\alpha V_\mathrm{n})(\nabla^\alpha V_\mathrm{n})$, which is nonzero for a rotation about an axis lying in the plane of the membrane. Thus we conclude that $\mu_1=0$.

The membrane is also subject to tractions from the viscous solvent. The stress tensor for a viscous fluid is 
\begin{equation}
\sigma_{ij}=-p\delta_{ij}+\eta(\bnabla_i u_j+\bnabla_j u_i).
\end{equation}
For membranes and interfaces, there is no systematic expansion that leads to simple approximations analogous to slender-body theory, although an approximation similar to resistive-force theory is sometimes used for illustrative purposes~\cite{CaiLubensky1995}. 
Balancing all the forces on a patch of membranes, we find
\begin{equation}
\nhat_i(\sigma^+_{ij}-\sigma^-_{ij})+\nabla_\alpha(\bF^\alpha_\mathrm{tot})_j=0,
\end{equation}
where $\sigma^\pm_{ij}$ is the three-dimensional viscous stress evaluated at the membrane surface with $\nhat$  the outward-pointing normal, and $\bF^\alpha_\mathrm{tot}$ is the elastic and viscous stresses of Eqs.~(\ref{Fabend}), (\ref{Fas}) and~(\ref{totmemvisstr}). 

We close this subsection by writing out the equations of motion for a membrane which is almost flat. We work to lowest order in deflection, and study the evolution of a ripple with wavelength $2\pi/q$. The membrane shape is give by $\br(x,y,t)=(x,y,z(x,y,t))$, where
$z(x,y,t)=h(t)\exp(\mathrm{i}q x)$,  and complex notation is used
since we work to linear order in $h$. The large value of the membrane bulk modulus implies that tangential flow in the membrane is small, which is why there is no motion in the $xy$-plane in the expression for $\br(x,y,t)$. 
The equations of motion for the membrane simplify greatly since the membrane flow vanishes, $\mathbf{V}=\mathbf{0}$, and many of the geometrical quantities such as $K$ are second order in $h$.  The mean curvature is 
$H\approx\nabla^2z/2$, where to leading order $\nabla^2\approx \partial^2/\partial x^2+\partial^2/\partial y^2$. The normal component of force balance on the membrane is thus
\begin{equation}
\left(\sigma_{zz}^+-\sigma_{zz}^-\right)_{z=0}-\kappa\nabla^4z=0\label{ftot}.
\end{equation}
To find the stresses $\sigma_z^\pm$, solve the Stokes equations~(\ref{stokes1}--\ref{stokes2}) subject to the no-slip boundary conditions $u_x=0$ and $u_z={\dot h}\exp(\mathrm{i}qx)$:
\begin{eqnarray}
u_x^\pm&=&-qz{\dot h}\mathrm{e}^{\mathrm{i}qx\mp qz}\\
u_z^\pm&=&{\dot h}(1\pm qz)\mathrm{e}^{\mathrm{i}qx\mp qz}\\
p^\pm&=&\pm2\eta q{\dot h}.
\end{eqnarray}
Thus, $[\sigma_{zz}^+-\sigma_{zz}^-]_{z=0}=-4\eta q{\dot h}\exp{\mathrm{i}qx}$, which together with Eq.~(\ref{ftot}) implies
\begin{equation}
{\dot h}=-\frac{\kappa q^3}{4\eta}h
\end{equation}
The ripple relaxes with the bending relaxation time~(\ref{bendrelaxtime})~\cite{BrochardLennon1975}.
Note that even if $q\ell_\mathrm{SD}\gg1$, the membrane viscosity does not enter since there are no tangential flows due to the high bulk modulus. However, curvature can lead to tangential flows for incompressible membranes, as we show in the next section.

\subsection{Illustrative example: buckling and pearling instabilities of tubular polymersomes}\label{sec:polymersomes}

In this section we discuss two instabilities of a tubular vesicle: pearling and buckling. Pearling arises when the membrane is subject to tension, and buckling arises when the membrane is under compression.  Both instabilities consist of a cylinder deforming into a series of bulges, but the physical mechanisms are different. Pearling can be induced by applying a laser tweezer to a tubular liposome~\cite{bar-ziv_moses1994,Bar-ZivMosesNelson1998}. The laser tweezer induces a tension, and if the tension is sufficiently high, the cylinder can lower its energy by deforming into series of bulges.  This instability is similar to the Rayleigh-Plateau instability of cylindrical interfaces~\cite{Rayleigh1892,Tomotika1935}, and it has been extensively studied in liposomes
\cite{nelson_powers_seifert1995,GranekOlami1995,goldstein_etal1996}. Buckling of a membrane is analogous to the buckling of a rod under tension~\cite{landau_lifshitz_elas}. Buckling of a spherical membrane is well-studied~\cite{DeulingHelfrich1976,DeulingHelfrich1976JdP,Jenkins1977,OYZhongCanHelfrich1987,Peterson1988,Peterson1989}, but the buckling of a cylindrical membrane has received less attention. 

Consider a cylindrical polymersome of radius $a$. If $a$ is of the order of microns, then $\ell_\mathrm{SD}/a\gg1$, and surface membrane viscosity cannot be disregarded. 
We will study the instability of the cylinder as a function of the pressure $p$ inside the vesicle, with the pressure outside taken to be zero. Increasing the pressure increases the tension in the membrane and leads to pearling; decreasing the pressure compresses the membrane and leads to buckling.  Suppose that in the initial state that there are just enough molecules to enclose a cylindrical volume of radius $a$ without stretching the membrane. Thus, $\phi=\phi_0$,  $\gamma=-p_{2\mathrm{d}}=0$, and there is no stretching stress in the membrane, $\bF^\alpha=\mathbf{0}$ [Eq.~(\ref{Fas})]. However, because the membrane is curved, there is a nonzero bending stress $\bF^\alpha_\mathrm{bend}$ and a corresponding bending force per unit area [Eq.~(\ref{fmembrane})], which must balance the internal pressure: $p\nhat+\beff_\mathrm{bend}=\mathbf{0}$. Since $H=-1/(2a)$, normal force balance in the initial equilibrium state yields $p=p_0=-\kappa/(2a^3)$.

Now imagine changing the pressure to a value $p\neq p_0$. The cylinder stretches or compresses, depending on whether $p>0$ or $p<0$. Since the stretch modulus $k$ is high, $ka^2/\kappa\gg1$, there is little stretch: $\phi_1/\phi_0=(\phi-\phi_0)/\phi_0\ll1$. However,  the tension $\gamma$ is \textit{not} small [cf. discussion after Eq.~(\ref{Fas})], and 
\begin{equation}
\gamma=p+\kappa/(2a^3).\label{gispk}
\end{equation}
Our task is to determine if sufficiently large (positive or negative) tension in the membrane destabilizes the cylindrical shape.

To proceed, parameterize the vesicle in cylindrical coordinates, $\br(\varphi,z,t)=\hat{\brho}(a+h(z,t))+(z+w(z,t))\hat\mathbf{z}$. Note that  the deformed shape is assumed axisymmetric. To first order in the displacements $h$ and $w$, the mean curvature is
$H\approx-(1/a+h/a^2+h'')/2$, and the Gaussian curvature is
$K\approx-h''/a$.
The evolution of the shape of the membrane is governed by normal force balance,
\begin{equation}
\nhat\cdot(\beff_\mathrm{vis}+\beff_\mathrm{bend}+\beff_\mathrm{s})+(\sigma_{nn}^+-\sigma_{nn}^-)_{\rho=a}=0,
\end{equation}
where $\sigma_{nn}^\pm|_{\rho=a}=\nhat_i\sigma_{ij}^\pm(\rho=a,z,t)\nhat_j$ is the normal-normal component of the stress tensor for the solvent outside ($+$) and inside ($-$) the vesicle evaluated at the membrane, to leading order in $h$.

We suppose the deformation is sinusoidal, $h(z,t)\propto \exp\mathrm{i}(qz)$. Again, since $ka^2/\kappa\gg1$, the change in density is small and we may use the condition of continuity for an incompressible membrane, Eq.~(\ref{VVn}), to relate the shape to the flow $V^z=\dot w$ in the membrane.  To linear order, 
\begin{equation}
\mathrm{i}qV^z=-\dot{h}/a.
\end{equation}
The internal forces per unit area acting on the membrane are thus
\begin{equation}
\nhat\cdot(\beff_\mathrm{vis}+\beff_\mathrm{bend}+\beff_\mathrm{s})\approx-2\mu \dot{h}/a^2-\mathcal{E}(q)h,
\end{equation}
where $\mathcal{E}(q)=\kappa(q^4a^2+\bar{p}q^2a^2+1-\bar{p})/a^4$, and we have introduced the dimensionless pressure $\bar{p}=pa^3/\kappa$.

\begin{figure}
\centerline{%\centerbmp{4.95 in}{2.84 in}{hyperact-fig.bmp}
\includegraphics[height=2.0in]{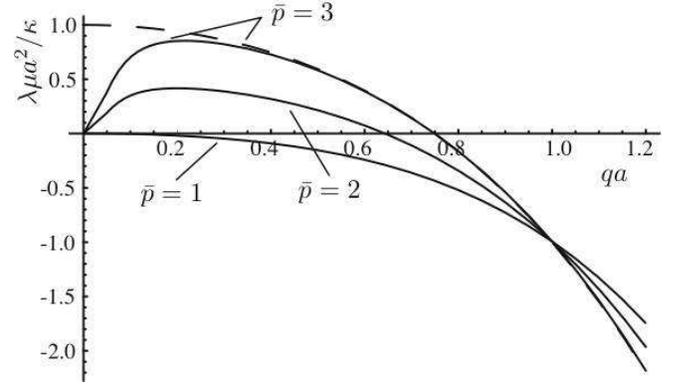}}
\caption{Dimensionless growth rate $\lambda\mu a^2/\kappa$ vs. dimensionless wavenumber $qa$ for the pearling instability for a tubular polymersome, in which membrane surface viscosity dominates the dissipation. The solid lines have $\ell_\mathrm{SD}/a=1000$ and the dashed line has $\ell_\mathrm{SD}/a=\infty$.}
\label{fig:pearling}
\end{figure}

The solvent forces acting on the membrane are determined by solving Stokes equations~(\ref{stokes1}--\ref{stokes2}) for the flow $\mathbf{u}^\pm$ inside and outside the tube,  and then calculating the stress $\sigma^\pm_{nn}$. Note that the imposed pressure $p$ enters the  stress tensor through the dynamic pressure $P$: $\sigma_{ij}^\pm=-P^\pm\delta_{ij}+\eta(\partial_i u^\pm_j+\partial_ju^\pm_i)$, where $P^-=p$ when $\dot{h}=0$, and $P^+=0$ when $\dot{h}=0$. Since the Stokes equations are linear, the stress at the membrane has the form
$(\sigma_{nn}^+-\sigma_{nn}^-)_{\rho=a}=-\eta\Lambda^{-1}\dot{h}+p$, where the ``dynamical factor" $\Lambda$ depends only on the wavenumber $q$ and radius $a$.
The Stokes equations in an axisymmetric geometry are readily solved using the stream function $\psi^\pm$, where $\mathbf{u}^\pm=\bnabla\times(\psi^\pm\hat\mathbf{z})$~\cite{happel_brenner1965}. Here we only quote the result and refer to~\cite{goldstein_etal1996} for the details. 
With the no-slip boundary conditions at the membrane,  $u_\rho^\pm(a)=\dot h$ and $u_z^\pm(a)=V_z$, the dynamical factor is
\begin{equation}
\Lambda=-\frac{a}{2}\frac{\left[qa(I_0^2-I_1^2)-2I_0I_1\right]\left[qa(K_0^2-K_1^2)+2K_0K_1\right]}{I_1K_1+I_0K_0+qa(I_1K_0-I_0K_1)},
\end{equation}
where $I_n$ and $K_n$ are modified Bessel functions of order $n$ evaluated at $qa$.
Note that this expression corrects an error in equation (D.9) of~\cite{goldstein_etal1996}, which gave a dynamical factor which is too small.

Balancing internal membrane forces and the solvent force leads to $\dot{h}=\lambda h$, where
\begin{equation}
\lambda=-\frac{\kappa}{\mu a^2}\frac{q^4a^4+ \bar{p} q^2a^2+1-\bar{p}}{2+(a/\Lambda)(a/\ell_\mathrm{SD})}.\label{polygrowth}
\end{equation}
The sign of $\lambda$ is determined by the sign of $\mathcal{E}(q)=\kappa(q^4a^4+\bar{p}q^2a^2+1-\bar{p})/a^4$. When $p>\kappa/a^3$, $\mathcal{E}>0$ and the cylinder is unstable.  By Eq.~(\ref{gispk}), this pressure corresponds to a tension $\gamma=3\kappa/(2a^3)$, the critical tension for the pearling instability~\cite{goldstein_etal1996}. Figure~\ref{fig:pearling} shows the growth rate $\lambda$ for the pearling instability for various pressures and $\ell_\mathrm{SD}/a=1000$. Also shown for comparison is the growth rate for $p=3\kappa/a^3$ and $\ell_\mathrm{SD}/a=\infty$. Note how $\Lambda$, which arises from the dynamics of the incompressible solvent, forces the growth rate to vanish at $q=0$, even when $\ell_\mathrm{SD}/a\gg1$.

\begin{figure}
\centerline{%\centerbmp{4.95 in}{2.84 in}{hyperact-fig.bmp}
\includegraphics[height=2.0in]{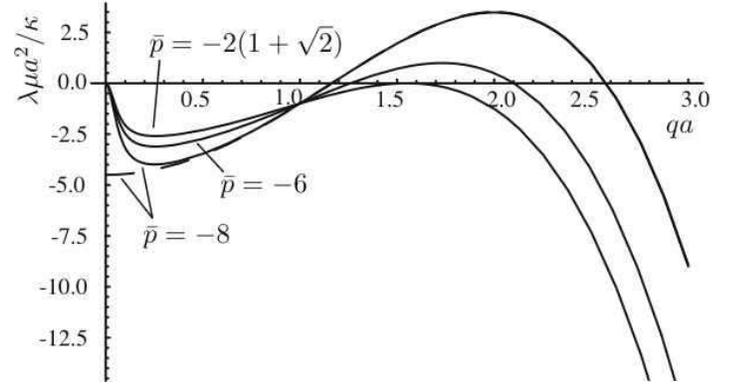}}
\caption{Dimensionless growth rate $\lambda\mu a^2/\kappa$ vs. dimensionless wavenumber $qa$ for the buckling instability for a tubular polymersome, in which membrane surface viscosity dominates the dissipation. The solid lines have $\ell_\mathrm{SD}/a=1000$ and the dashed line has $\ell_\mathrm{SD}/a=\infty$.}
\label{fig:buckling}
\end{figure}

When $p<-2(1+\sqrt{2})\kappa/a^3\approx-4.828\kappa/a^3$, the cylinder is also unstable. Since negative pressure corresponds to compressive stress, this instability is analogous to the Euler buckling instability of a rod. 
Figure~\ref{fig:buckling} shows the growth rate for the buckling instability. Note the contrast with the pearling instability. In the buckling instability, when the pressure is just sufficient to induce the instability, only a narrow band of wavenumber centered about a nonzero wavenumber [$q^*a=(1+\sqrt{2})$] has positive growth rate. In the pearling instability, the band extends from $q=0$ to a finite value, and therefore there is no limit to the wavelength of the unstable modes.

\section{OUTLOOK}
The aim of this review has been to give a practical overview of the geometrical tools necessary to formulate the equations of motion of polymers and membranes. Equilibrium and dynamic properties of thin filaments and membranes are best described using geometrical ideas such as curvature. We have also seen how naturally these geometric ideas fit with the variational point of view. In the case of filaments, the variation of the energy led directly to the constitutive relation between moments and the rate of rotation of the directors, or strain. In the case of membranes, variation using auxiliary variables provided a quick route to the elastic forces per area, with the added bonus of the explicit form of the stress tensor of the membrane. Geometric ideas are also crucial for understanding viscous flow on a deforming curved surface.
This study of polymers and membranes remains an active area of research, and the purely mechanical problems considered here provide a solid foundation for probing the dynamics of fluctuating polymers and membranes. 

\section*{Acknowledgments}

I would like to thank  Ray Goldstein, Greg Huber, Randy Kamien, Phil Nelson, and Charles Wolgemuth for many discussions. I am grateful to Charles Wolgemuth for the code that produced Fig.~\ref{circlehelix}, and Deborah Fygenson for Fig.~\ref{tubule}. Some of this review was written at the Aspen Center for Physics.
This work was supported in part by National Science Foundation Grants Nos. NIRT-0404031, DMS-0615919, and CBET-0854108.

\appendix\section{Coordinate transformations}\label{coordtmns}

In this appendix we describe how the components of tensors transform under a change of coordinates. In general, a tensor is a quantity that transforms linearly under a change of coordinates. We begin with the metric tensor as a concrete example.

Consider two systems of coordinates related by
 $\bar\xi^\alpha=\bar\xi^\alpha(\xi^1,\xi^2)$. From the chain rule,  the metric tensor $\bar g_{\alpha\beta}$ in the new coordinate system is related to metric in the old coordinates by a linear transformation law:
\begin{equation}
\bar g_{\alpha\beta}=\frac{\partial\br}{\partial\bar\xi^\alpha}\cdot\frac{\partial\br}{\partial\bar\xi^\beta}=\frac{\partial\xi^\gamma}{\partial\bar\xi^\alpha}\frac{\partial\xi^\delta}{\partial\bar\xi^\beta}g_{\gamma\delta}.\label{gcovariant}
\end{equation}
The differentials have the opposite transformation law,
\begin{equation}
\dd \bar\xi^\alpha = \frac{\partial\bar\xi^\alpha}{\partial\xi^\beta}\dd\xi^\beta,
\end{equation}
which ensures that
\begin{equation}
g_{\alpha\beta}\dd\xi^\alpha\dd\xi^\beta=\bar g_{\alpha\beta}\dd\bar\xi^\alpha\dd\bar\xi^\beta.
\end{equation}
Quantities that transform linearly under a change of coordinates are tensors. 
The tensor transformation law for $g_{\alpha\beta}$ reflects the fact that the distance $\dd s^2$ is a geometric quantity, independent of the choice we make for coordinates.

It is traditional to use upper or lower indices to indicate what kind of transformation law a tensor has.
Thus, a tensor $T^{\alpha_1\alpha_2\dots}_{\beta_1\beta_2\dots}$  transforms homogeneously under a change of coordinates as
\begin{equation}
\bar T^{{\alpha}_1\alpha_2\dots}_{\beta_1\beta_2\dots}=\frac{\partial\bar\xi^{\alpha_1}}{\partial\xi^{\gamma_1}}\frac{\partial\bar\xi^{\alpha_2}}{\partial\xi^{\gamma_2}}\cdots
\frac{\partial\xi^{\delta_1}}{\partial\bar\xi^{\beta_1}}\frac{\partial\xi^{\delta_2}}{\bar\partial\xi^{\beta_2}}\cdots T^{{\gamma}_1\gamma_2\dots}_{\delta_1\delta_2\dots}.
\end{equation}
Tensors with upper indices are ``contravariant," and tensors with lower indices are ``covariant." 

As mentioned in \S\ref{sec:surfaces}, we use raised indices to denote  the inverse of the metric tensor:
\begin{equation}
g^{\alpha\gamma}g_{\gamma\beta}=\delta^\alpha_{\ \beta}.\label{g-ginv}
\end{equation}
Viewing Eq.~(\ref{gcovariant}) as a matrix equation and inverting yields 
\begin{equation}
\bar g^{\alpha\beta}=\frac{\partial\bar\xi^\alpha}{\partial\xi^\gamma}\frac{\partial\bar\xi^\beta}{\partial\xi^\delta}g^{\gamma\delta};\label{ginv-covariant}
\end{equation}
that is,  $g^{\alpha\beta}$  transforms as a contravariant tensor. By using the metric tensor, we can create a covariant tensor from a contravariant one, and vice-versa. For example, $W_\alpha=g_{\alpha\beta}W^\beta$ and $W^\alpha=g^{\alpha\beta}W_\beta$. Geometrically, we can think of $W_\alpha$ as the components of the vector field $\mathbf{W}=W^\alpha\bt_\alpha$ with respect to a dual basis $\bt^\alpha$ defined by
\begin{equation}
\bt^\alpha\cdot\bt_\beta=\delta^\alpha_{\ \beta}.
\end{equation}
In other words, $\mathbf{W}=W^\alpha\bt_\alpha=W_\alpha\bt^\alpha$.
The dual basis vectors $\bt^\alpha$ are like reciprocal lattice vectors, and also obey the rule about raising and lowering indices:
\begin{equation}
\bt^\alpha=g^{\alpha\beta}\bt_\beta.
\end{equation}

Recall that the covariant derivative was defined in \S\ref{sec:surfaces} to be a geometric object, independent of coordinates once the direction $\bt_\alpha$ is chosen. The quantity $(\nabla_\alpha V)^\beta=\bt^\beta\cdot\partial_\alpha\mathbf{V}$ is a tensor since $\partial_\alpha\mathbf{V}$ and $\that^\beta$ are both tensors. Note however that the Christoffel symbol $\Gamma^\gamma_{\alpha\beta}$ is not a tensor.

The importance of these linear transformation laws is that they allow us to easily construct coordinate-invariant expressions. A quantity is coordinate invariant when  there is an equal number of upper and lower indices that are summed over. For example, the mean curvature $H=K^{\alpha}_{\ \alpha}$ and the divergence $\nabla_\alpha V^\alpha$  are both invariant under coordinate changes. Likewise, the Gaussian curvature $\det K^\alpha_{\ \beta}$ is invariant under coordinate changes, since the upper and lower indices have associated with them transformation matrices that are inverses.
On the other hand, $g=\det g_{\alpha\beta}$ is not coordinate-invariant since it has two lower indices and no upper indices, and therefore gets two factors of the Jacobian matrix $|\partial(\xi^1,\xi^2)/\partial(\bar\xi^1,\bar\xi^2)|$ under a coordinate change. These factors are precisely what is needed to make $\mathrm{d}S=\sqrt{g}\mathrm{d}\xi^1\mathrm{d}\xi^2$ invariant under coordinate change. 

\section{Green's theorem}\label{green}

Consider Stokes theorem for a tangent vector field $\mathbf{A}=A^\alpha\bt_\alpha$ on a region $D$ of a surface,
\begin{equation}
\int_D \nhat\cdot\bnabla\times\mathbf{A}\dd S=\oint_{\partial D} \mathbf{A}\cdot\dd\mathbf{s},
\end{equation}
where $\dd\mathbf{s}$ is the infinitesimal tangent vector along boundary, and $\bnabla$ denotes the three-dimensional gradient in Cartesian coordinates. Note that $\bnabla$ is the covariant derivative for Euclidean space in standard coordinates.  Recall that the normal component of the curl has derivatives along the tangent direction only:
\begin{eqnarray}
\nhat\cdot\bnabla\times{\mathbf A}&=&(\nhat\times\bnabla)\cdot\mathbf{A}\nonumber\\
&=&\left[\frac{\bt_1\times\bt_2}{\sqrt{g}}\times\bnabla\right]\cdot\mathbf{A}\nonumber\\
&=&\frac{1}{\sqrt{g}}\left[\bt_2( \bt_1\cdot\bnabla )- \bt_1 (\bt_2\cdot\bnabla)\right]\cdot A^\alpha\bt_\alpha\nonumber\\
&=&\frac{1}{\sqrt{g}}(\nabla_1 A_2-\nabla_2 A_1)\nonumber\\
&=&\frac{1}{\sqrt{g}}(\partial_1 A_2-\partial_2 A_1)
\end{eqnarray}
Since $\mathbf{A}\cdot\dd\mathbf{s}=A_\alpha\dd\xi^\alpha$, we have shown that
\begin{equation}
\int \epsilon^{\alpha\beta}\nabla_\alpha A_\beta\dd S=\oint A_\alpha\dd\xi^\alpha, 
\end{equation}
where we have introduce the two-dimensional antisymmetric tensor with components $\epsilon^{12}=-\epsilon^{21}=1/\sqrt{g}$ and $\epsilon^{11}=\epsilon^{22}=0$. Defining $\epsilon_{\alpha\beta}=g_{\alpha\gamma}g_{\beta\delta}\epsilon^{\gamma\delta}$ (note $\epsilon_{12}=\sqrt{g}$) and $A_\alpha=\epsilon_{\beta\alpha}B^\beta$, we also have
\begin{equation}
\int\nabla_\alpha B^\alpha\dd S=\oint\epsilon_{\beta\alpha}B^\beta\dd\xi^\alpha.
\end{equation}
A short calculation shows that $\epsilon_{\alpha\beta}B^\alpha\dd\xi^\beta=\mathbf{B}\cdot\hat\mathbf{p}\dd s$, where $\mathbf{B}=B^\alpha\bt_\alpha$; $\hat\mathbf{p}$ is the outward-pointing unit vector normal to the curve $\partial D$ defining the region of integration and lying in the tangent plane, $\hat\mathbf{p}=\bt_\alpha\dd\xi^\alpha\times\nhat/\sqrt{g_{\beta\gamma}\dd\xi^\beta\dd\xi^\gamma}=\partial\mathbf{r}/\partial s\times\nhat$; and $\dd s=\sqrt{g_{\beta\gamma}\dd\xi^\beta\dd\xi^\gamma}$ is measure of arclength along the curve. Therefore we may write
\begin{equation}
\int_D\nabla_\alpha B^\alpha\dd S=\oint_{\partial D} B^\alpha p_\alpha\dd s
\end{equation}
where $p_\alpha\dd s=\epsilon_{\alpha\beta}\dd\xi^\beta$.

%\bibliographystyle{apsrmp}
%\bibliography{../../../newrefs}
%\bibliography{../../../newrefs,../../LowReReview/review}
%\newpage

\end{document}